\documentclass[manuscript]{aastex}
\usepackage{lscape}
\pdfoutput=1

\slugcomment{Accepted for publication in the Astrophysical Journal}

\shorttitle{Super-damped Ly-$\alpha$ Absorbers}
\shortauthors{Kulkarni et al.}

\begin{document}

\title{Keck and VLT Observations of Super-damped Lyman-alpha Absorbers at $z \sim2- 2.5$: Constraints on Chemical Compositions and Physical Conditions
\footnote { 
 Includes observations collected during program 
ESO 93.A-0422 at the European Southern Observatory (ESO) Very Large
Telescope (VLT) with the Ultraviolet-Visual Echelle Spectrograph  (UVES) on the 8.2 m telescopes operated at
the Paranal Observatory, Chile.  Some of the data presented herein were obtained at the W.M. Keck 
Observatory, which is operated as a scientific partnership among the California Institute of Technology, the University of California and the National Aeronautics and 
Space Administration. The Observatory was made possible by the generous financial support of the W.M. Keck Foundation.  }}

\author{Varsha P. Kulkarni}
\affil{University of South Carolina, Dept. of Physics and Astronomy, Columbia, SC 29208}
\email{kulkarni@sc.edu}

\author{Debopam Som}
\affil{University of South Carolina, Dept. of Physics and Astronomy, Columbia, SC 29208;  
Aix Marseille Universit\'e, CNRS, Laboratoire dÕAstrophysique de Marseille, UMR 7326, 13388, Marseille, France}

\author{Sean Morrison}
\affil{University of South Carolina, Dept. of Physics and Astronomy, Columbia, SC 29208}

\author{Celine P\'eroux and Samuel Quiret}
\affil{Aix Marseille Universit\'e, CNRS, Laboratoire d'Astrophysique de Marseille, UMR 7326, 13388, Marseille, France}

\author{Donald G. York}
\affil{Department of Astronomy \& Astrophysics, University of Chicago, Chicago, IL 60637}

\begin{abstract}
We report Keck/ESI and VLT/UVES observations of three super-damped Lyman-$\alpha$ quasar absorbers with H I 
column densities log $N_{\rm H I}$$\ge$21.7 at redshifts 2$\la$z$\la$2.5. All three absorbers show similar metallicities ($\sim$-1.3 to -1.5 dex), and dust depletion 
of Fe, Ni, and Mn. Two of the absorbers show supersolar [S/Zn] and [Si/Zn]. We combine our results with those for other DLAs 
to examine trends between $N_{\rm H I}$, metallicity, dust depletion.  A larger fraction of the super-DLAs lie close to or above the line [X/H]=20.59$-$log $N_{\rm H I}$ in the metallicity vs.  $N_{\rm H I}$ plot, compared to the less gas-rich DLAs, suggesting that super-DLAs are more 
likely to be 
rich in molecules. Unfortunately, 
our data for Q0230-0334 and Q0743+1421 do not  cover H$_{2}$ absorption lines. 
For Q1418+0718, some H$_{2}$ lines are covered, but not detected. CO is not detected 
in any of our absorbers. For DLAs with log $N_{\rm H I}$$<$21.7, we confirm strong correlation between metallicity and Fe depletion, and find a correlation   
between metallicity and Si depletion. For super-DLAs, these correlations are weaker or absent. The absorbers toward Q0230-0334 
and Q1418+0718 show potential detections of weak Ly-$\alpha$ emission, implying star formation 
rates of  $\sim$1.6 and  $\sim$0.7 M$_{\odot}$ yr$^{-1}$, respectively (ignoring dust extinction). 
Upper limits on the electron densities from  C~II$^{*}$/C~II or Si~II$^{*}$/Si~II are low, but are higher than the median values in less gas-rich DLAs. Finally, systems with 
log $N_{\rm H I} > 21.7$ may have somewhat narrower velocity dispersions $\Delta v_{90}$  than the less gas-rich DLAs, 
and may arise in cooler and/or less turbulent gas. 
\end{abstract}


\section{Introduction}
Galaxies such as the Milky Way contain a large fraction of their present-day mass in stars. However, their 
chemically less-enriched analogs at higher redshifts are more dominated by gas. How galaxies exchange gas 
with their surroundings, and how they progressively convert this gas into stars and enrich it chemically remain as 
some of the core questions in astrophysics. 

Star formation is well-known to be associated with cold, neutral gas. Indeed, the star formation rate (SFR) is known to be strongly 
correlated with the surface density of neutral gas, as per the Kennicutt-Schmidt law (e.g., Schmidt 1959; Kennicutt 1998). The most gas-rich galaxies thus offer an 
opportunity to study the most vigorously star-forming regions. Furthermore, such regions can potentially allow a look at interesting  
chemical enrichment processes in young galaxies. 

A powerful technique to study gas-rich galaxies is to search for their strong absorption signatures in the light of background sources 
such as quasars or gamma-ray bursts (GRBs). Damped Lyman-alpha absorbers (DLAs) with neutral hydrogen column densities 
$N_{\rm H I} \ge 2 \times 10^{20}$ cm$^{-2}$ and sub-DLAs with $10^{19} \le N_{\rm H I} < 2 \times 10^{20}$
cm$^{-2}$ dominate the mass density of neutral gas in the universe (e.g., P\'eroux et al. 2005; 
Prochaska \& Wolfe 2009; Noterdaeme et al. 2012a; Zafar et al. 2013), and may be the progenitors of galaxy
disks or halos. Absorbers with log $N_{\rm H I} \ge 21.7$ have been detected in many GRBs ($> 60 \%$ of GRB-DLAs), but are very rare ($\sim$0.5$\%$) 
among quasar DLAs. Despite their small numbers, these extremely gas-rich systems may contribute $\sim$10$\%$ of the comoving 
neutral gas density (Noterdaeme et
al. 2012a, 2014). About 100 such ``super-DLAs" have been discovered in the SDSS spectra. Study of these extremely gas-rich systems 
offers an excellent opportunity to understand star formation and chemical enrichment in young, gas-rich galaxies.  Such studies are 
especially important at redshifts $2 < z < 3$, corresponding to the epoch of peak global star formation activity (e.g., 
Madau et al. 1998; Bouwens et al. 2011). 

The first detailed studies of an intervening quasar super-DLA were those of the $z=2.2$ absorber toward SDSSJ1135-0010, carried out by  
Kulkarni et al. (2012) and Noterdaeme et al. (2012b) using the Very Large Telescope (VLT) Ultraviolet-Visual Echelle Spectrograph (UVES) and
 X-shooter. This super-DLA is fairly enriched,  with 
[Zn/H] = $-1.06 \pm 0.10$, [Si/H] = $-1.10 \pm 0.10$, [Cr/H] = $-1.55 \pm 0.10$, [Ni/H] = $-1.60 \pm 0.10$, [Fe/H] = $-1.76\pm 0.10$, [Ti/H] =
$-1.69 \pm 0.11$, [P/H] = $-0.93 \pm 0.23$, and [Cu/H] = $-0.75 \pm 0.14$. Furthermore, it shows strong 
Ly-$\alpha$ emission near the bottom of the DLA trough and other nebular emission lines, implying a fairly high SFR of $\sim 25$ M$_{\odot}$ yr$^{-1}$.   The Ly-$\alpha$ emission shows two 
distinct peaks (each $>$ 7$\sigma$), possibly suggesting outflowing gas (Noterdaeme et al. 2012b).

In fact, stacked SDSS spectra of quasars with Super-DLA absorbers (shifted to the rest-frame of the absorbers) show a
statistical detection of Ly-$\alpha$ emission at the bottom of the DLA trough, suggesting close association with Ly-$\alpha$ emitters 
(Noterdaeme et al. 2014) and small impact parameter (so that the emission falls in the fiber). There are also indications that at least some super-DLAs may be rich in molecular gas (e.g., Guimaraes et al. 2012; 
Noterdaeme et al. 2015b). 
Such quasar super-DLAs may resemble
the DLAs arising in GRB hosts (e.g., Guimaraes et al. 2012). Indeed, GRB DLAs possibly arise in inner parts of galaxies, while most (less gas-rich) quasar DLAs probably arise in outer parts. 

In principle, large H I column densities could also be associated with large sizes and hence large masses of absorbing galaxies. It is therefore also interesting to 
examine whether super-DLAs show larger velocity dispersions, using measurements of metal absorption line profiles. 

With the goal of further understanding the unique properties of this rare class of galaxies, we have recently started obtaining follow-up 
spectroscopy of quasars that show super-DLA absorbers in SDSS spectra. Here we report a study of 3 super-DLAs 
at $2 \lesssim z \lesssim 2.5$, 
based on our observations obtained with the Keck and the VLT. 

This paper is organized as follows: Section 2 describes the observations, data reduction, and absorption line measurements. Section 3 presents the results of profile fitting and  column 
density measurements. Section 4 discusses the results and compares the chemical compositions and physical properties of super-DLAs with those for 
other DLAs presented in the literature. Section 5 summarizes our main conclusions. 

\section{Observations and Data Reduction}

Table 1 lists the targets. These quasars were selected for our study because their SDSS spectra show damped Lyman-alpha absorbers 
with log $N_{\rm H I} \ge 21.7$ and the presence of several metal lines (Noterdaeme et al.  2014). We obtained 
follow-up Keck ESI or VLT UVES spectra of these quasars to achieve higher spectral resolution than the SDSS data, 
which is essential for our goal of accurate element abundance measurements. 

\subsection{Keck ESI Observations}

The quasars Q0230-0334 and Q0743+1421 were observed in classical mode with the Keck Echellette Spectrograph and Imager (ESI; Sheinis et al. 2002) on November 7-8,  2013 under 
NOAO program 2013B-0525 (PI: V. Kulkarni). Severe weather problems resulted in a loss of 1.8 nights out of the 2 nights awarded, 
allowing us to obtain only one exposure for Q0230-0334 and three exposures for Q0743+1421.  

The ESI spectra were reduced and extracted using ESIRedux, an IDL-based reduction package written by Jason X. Prochaska. 
The extracted spectra from individual orders were split on average into 3 pieces (typically $\sim 100 - 400$ {\AA} wide), 
and these pieces were continuum-fitted using the IRAF ``CONTINUUM'' task. To fit the continuum in each piece, we tried both cubic spline 
and Legendre polynomials, typically of order $3-5$, and used the function which provided the best fit as judged from the 
RMS of the residuals. For the absorber towards J0743+1241, the continuum-normalized  pieces from the 3 exposures were  
combined into a single piece for each wavelength range using the IRAF ``SCOMBINE'' task. 

\subsection{VLT UVES Observations}

The quasar Q1418+0718 was observed with the UVES (Dekker et al. 2000) on VLT under program 
093.A-0422 (PI: C. P\'eroux) in Service Mode on May 26, June
18, and July 24, 2014. The object was observed using a
combined 437$+$760 nm setting with three different observations with
exposure times lasting 5400 sec each. The data were reduced using the
most recent version of the UVES pipeline in MIDAS (uves/5.4.3). Master
bias and flat images were constructed using calibration frames taken
closest in time to the science frames. The science frames were
extracted with the ``optimal" option. The resulting spectra were
combined, weighting each spectrum by the signal-to-noise ratio and
correcting to the vacuum heliocentric reference. The quasar continuum was fitted using the IRAF CONTINUUM task on regions of reasonable size (typically 
$\sim 100-300$ {\AA} wide). Cubic spline, Legendre, and Chebyshev polynomials of various orders were tried, and usually the cubic 
spline fits were found to be the best. 

\subsection{Absorption Line Measurements}

Column densities were determined 
by Voigt profile fitting using the program VPFIT\footnote{http://www.ast.cam.ac.uk/ rfc/vpfit.htm} version 10.0. 
Figs. 1, 2, and 3 show the Voigt profile fits to the H I Ly-$\alpha$ lines for the super-DLAs toward Q0230-0334, Q0743+1421, and Q1418+0718, respectively. 
For Q0743+1241, the ESI spectra did not cover the Ly-alpha absorption line of the super-DLA absorber; therefore the relevant region of the SDSS DR12 BOSS spectra was  fitted to 
measure the $N_{\rm HI} $ for this super-DLA,  using a 7th order Chebyshev function for continuum fitting (Fig. 2).
Figs. 4-5, 6-8, and 9 correspondingly show the profile fits to key metal lines in these systems. 

The total 
column densities were obtained by summing over all the velocity components. These values were checked 
independently using the apparent optical depth method  (AOD; see Savage \& Sembach 1991) for unsaturated, unblended lines. 
For consistency with past studies (which we use to compare to our results in section 4 below), we have adopted oscillator strengths 
from Morton et al. (2003). We note, however, that more recent oscillator strength determinations exist for some elements (e.g., 
Kisielius et al. 2014, 2015), and 
should be uniformly applied to all future element abundance studies in DLA/sub-DLAs. (For the key elements S and Zn, these revised oscillator strengths would 
result in DLA abundances higher by 0.04 dex and lower by 0.10 dex, respectively). Solar system abundances were adopted from 
Asplund et al. (2009). 

\section{ Results of Profile Fitting and Column Density  Measurements}

Tables 2 and 3  list the measurements of metal column densities for the absorber toward Q0230-0334. 
Tables 6 and 7  list the metal column density measurements for the absorber toward Q0743+1421, and 
Table 10 lists the metal column density measurements for the absorber toward Q1418+0718.  Tables 4, 8, and 11 give the total column densities 
(summed over individual velocity components) derived from the profile fits, along with the AOD estimates, if available. Table 5, 9 and 12 list the 
corresponding element abundances,   
calculated using the total metal column densities along with the H I column density, and using the solar abundances from Asplund et al. (2009). 
No ionization corrections were calculated, since the corrections are negligible at such high H I column densities. 

\section{Discussion}

\subsection{Element Abundances and Dust Depletions}

The element Zn is often used as a metallicity indicator in DLAs (e.g., Pettini et al. 1990, 1997). While there are several possible nucleosynthetic channels for Zn production (see, 
e.g., Matteucci et al. 1993; Kobayashi et al. 2006) 
making the nucleosynthetic origin of Zn 
unclear, on average Zn is observed to track Fe in the Milky Way halo stars, with 
[Zn/Fe] $\sim 0.0$ for  $0 < $ [Fe/H] $< -2.0$ and [Zn/Fe] $\sim 0.1$ dex for  -2.7 $<$ [Fe/H] $< $-2.0 (e.g., Mishenina et al. 2002; Nissen et al. 2004, 2007). Furthermore, Zn is relatively undepleted 
on interstellar dust grains (e.g., Jenkins 2009); thus the gas-phase abundance of Zn is a good indicator of the total Zn abundance, without the need for uncertain dust depletion corrections.  
By contrast, elements such as Fe, Cr, and Ni are significantly depleted on interstellar dust grains in the Milky Way. Therefore, their abundance relative to Zn gives an indication of dust depletion. 

All three super-DLAs in our sample show Zn abundances in the narrow range of -1.35 to -1.47 dex. 
Furthermore, we see [Fe/Zn] 
of -0.30, -0.59, and -0.48 dex, respectively, in the super-DLAs toward Q0230-0334, Q0743+1421, and Q1418+0718. Likewise, [Ni/Zn] in the three 
systems is  observed to be -0.29, -0.59, and -0.59 dex. Mn is even less abundant relative to Zn, with [Mn/Zn] of -0.39,  -0.74, and -0.81, respectively 
in  the absorbers toward Q0230-0334, Q0743+1421, and Q1418+0718.  
Cr appears to be weakly depleted in the absorbers toward Q0230-0334 and Q1418+0718, with [Cr/Zn] of  0.0 $\pm$ 0.11, and -0.13$ \pm$ 0.22, respectively. The absorber toward Q0743+1421 shows a 
higher Cr depletion, with [Cr/Zn] of  -0.41 $\pm$ 0.04. The low depletion of Cr seems surprising for the absorber toward Q0230-0334. 

Another  potentially interesting feature of the relative abundance patterns is that in two out of the three super-DLAs studied here, Si is observed to be more abundant 
relative to Zn than in the Sun,  with [Si/Zn] of  0.25 $\pm$ 0.10 dex and 0.23 $\pm$ 0.22 dex relative to the Sun. S, which shows nearly zero depletion in the Galactic interstellar medium (ISM; e.g. Savage \& Sembach 1996), is a better indicator of the abundance of the $\alpha$ elements than Si (which is depleted by $\sim 0.3-0.4$ dex 
even in the warm Galactic ISM e.g., Savage \& Sembach 1996; Jenkins 2009). It is 
therefore interesting to note that, in both systems for which we have S measurements, S is also more abundant relative to Zn than in the Sun. In the super-DLAs toward Q0230-0334 and Q1418+0718, the S/Zn ratio is 2.85 dex and 3.05 dex, respectively. 
Compared to the solar ratios, the S/Zn ratios in these super-DLAs are higher by 0.29 $\pm$ 0.20 and 0.49 $\pm$ 0.27 dex, respectively. S/Si is somewhat higher than solar: the S/Si ratios in the super-DLAs toward 
Q0230-0334 and Q1418+0718 are  -0.35 dex and 0.13 dex, respectively, i.e. higher by 0.04 $\pm$ 0.12 dex and 0.52 $\pm$ 0.17 dex respectively, than in the Sun. Of course, higher S/N observations are clearly needed to obtain smaller 
uncertainties in these relative abundance measurements). If Zn tracks Fe-peak elements in these absorbers, this could suggest that these systems have enhanced 
$\alpha$/Fe-group abundances (as seen, e.g., in the halo stars of the Milky Way), along with some Si depletion. We note, however, that Nissen \& Schuster (2011) found [Zn/Fe] $\sim 0.15$ dex for 
a subset of halo stars with high [$\alpha$/Fe], and  [Zn/Fe] $\sim 0.0$ dex for  halo stars with low [$\alpha$/Fe], and have suggested that the behavior of Zn is similar to that of $\alpha$ elements. 
However, in the absence of knowledge about the intrinsic $\alpha$/Fe ratio in DLAs (which is complicated by 
the higher depletion of Fe), it is not clear whether DLAs are similar to the high-$\alpha$ or low-$\alpha$ halo stars.  The large [S/Zn] and [Si/Zn] values in the super-DLAs toward Q0230-0334 and Q1418+0718 
suggest that Zn does not behave like an $\alpha$ element in these absorbers.  In any case, the large [S/Fe] values (0.59 dex and 0.97 dex) in the super-DLAs toward Q0230-0334 and Q1418+0718 
also suggest that they may show $\alpha$-enhancement (along with some dust depletion). 

\subsection{Search for Rare Metals}

Super-DLAs offer an excellent opportunity to study abundances of rare elements. The super-DLA toward Q1135-0010 showed the first detection of Cu 
in a quasar DLA, and in fact, with Cu/Zn exceeding that in the Sun, [Cu/Zn] = 0.31$ \pm$0.10 (Kulkarni et al. 2012). Such a level of [Cu/Zn] is surprising, 
given that Cu is depleted compared to Zn in the Galactic ISM (e.g., [Cu/Zn] $\sim$ -0.7 dex in the warm neutral medium  and $\sim$ -0.9 dex in the 
cold neutral medium  (Jenkins 2009).  
Observations of such rare elements can offer additional constraints on nucleosynthetic processes in the underlying galaxies. For example, V and Co are believed to be produced mainly in explosive Si burning in type II supernovae and to a smaller extent in 
type Ia supernovae (e.g., Woosley \& Weaver 1995; Bravo \& Martinez-Pinedo 2012; Battistini et al. 2015). 

For the super-DLAs toward Q0230-0334, Q0743+1421, and Q1418+0718, we searched the spectra for absorption lines of rare elements such as B, V, Co, Cu, Ge, and Ga, but no lines from these elements were 
detected. The 3 $\sigma$ upper limits on the column densities  
of these elements are listed under the ``AOD'' column  in Tables 4, 8, 11, and the corresponding abundance limits are included in Tables 5, 9, 12. The absorber toward Q0230-0334 shows under-abundant Cu, with 
[Cu/Zn] $<$-0.25; 
for the absorbers toward Q0743+1421 and Q1418+0718, the limits on Cu are less constraining with [Cu/Zn] $<$0.21 and $<$0.32, respectively. Both absorbers for which we have Co measurements show [Co/Zn] $<$-0.15. Higher S/N observations are necessary to obtain more definitive abundances of these rare elements. 

\subsection{Comparison with Other DLAs}

We now compare some of the key properties of the super-DLAs with those of other DLAs. (Throughout this discussion, we refer to the other DLAs with $20.3 \le$ log $N_{\rm H I} < 21.7$ as just ``DLAs'' or 
``moderate DLAs"). To construct the super-DLA sample, we combine our results with those for 11 other 
super-DLAs with $z_{abs} < z_{em}$,  that have element abundance determinations based on moderate or high-resolution spectroscopy (typically $R$ $>$ 6000). These other super-DLAs were reported by 
Prochaska et al. (2003), Heinmuller et al. (2006), Noterdaeme et al. (2007, 2008, 2015a), Kulkarni et al. (2012), Guimaraes et al. (2012), Ellison et al. (2012), and Berg et al. (2015). 

Observations show a deficit of the general population of DLAs with
high $N_{\rm H I}$ and high metallicity. Boisse et al. (1998) suggested that this deficit could arise from a dust selection effect: high-$N_{\rm H I}$, high-metallicity systems 
could be more dusty, and could 
obscure the background quasars more, thus remaining systematically under-represented in the high-resolution spectroscopic studies. 
On the other hand, the deficit of high-$N_{\rm H I}$, high-metallicity DLAs may arise from hydrogen becoming predominantly molecular, and hence 
undetectable in Ly-$\alpha$, above some $N_{\rm H I}$ threshold
(that decreases with increasing metallicity; Schaye 2001; Krumholz et al. 2009a). 

Fig. 10 compares the metallicity vs. H I column density data for DLAs and super-DLAs with detections of Zn or S toward quasars and GRB afterglows . The red filled circles and blue unfilled 
triangles show the measurements for quasar super-DLAs from our work and those from the literature.  
The black unfilled squares denote the measurements for other quasar DLAs with log $N_{\rm H I} < 21.7$ (from many references such as Ledoux et al. 2006, 
P\'eroux et al. 2006, Meiring et al. 2006, Prochaska et al. 2001, 2007, Noterdaeme et al. 2008, 
Rafelski et al. 2012, 2014; see Kulkarni et al. 2007, 2010 and Som et al. 2015 for further details). The orange and green diamonds denote the 
measurements for moderate DLAs and super-DLAs toward GRB afterglows. The
GRB data are from Vreeswijk et al. (2004), Prochaska et al. (2007), Ledoux et al. (2009), Savaglio et al. (2012), Cucchiara et al. (2015), and references therein. The GRB DLAs and super-DLAs are 
at the GRB redshifts and are thus associated with the GRB host galaxies.  [We note that, while we have used the $N_{\rm H I}$ reported in these studies for the GRB DLAs, the H I behind the GRB is not 
sampled by the GRB DLA sightlines, since the GRB is located within the host galaxy. Thus the true $N_{\rm H I}$ for the GRB sightlines may be expected to be higher by a factor of 2 (i.e., 0.3 dex) on average. 
The total metal columns along those sightlines may also be proportionately larger, and so the metallicities may not be that different.]  We show in solid orange  the line [X/H]=20.59$-$log $N_{\rm H I}$ corresponding to the 
``obscuration threshold''  that was suggested by Boisse et al. (1998) as a potential way to explain the deficit of DLAs with large $N_{\rm H  I}$ and high metallicity. 

The short-dashed, dotted, dot-dashed, and long-dashed lines in Fig. 10 show the curves calculated by Krumholz et al. (2009a) for ``covering'' fractions  $c_{\rm {H_{2}}}$ = 0.01, 0.05, 0.5, 
and 1.0 for the cross-section of the spherical molecular core of the cloud (surrounded by a shell of atomic gas).  These $c_{\rm H_{2}}$ values 
correspond to molecular mass fractions $f_{\rm H_{2}} = M_{\rm H_{2}}/M_{\rm total} = [1 -  \{(1 - c_{\rm H_{2}}^{-1.5})/ \phi_{mol} \}]^{-1} = 0.010$, 
0.102, 0.845, and 1.0, respectively. [The quantity $\phi_{mol}$ is the ratio of molecular gas density to atomic gas density, and is about 10 
(Krumholz, McKee, \& Tumlinson 2009b)]. 

Several interesting facts emerge from Fig. 10: (1) A much larger fraction of the quasar super-DLAs ($\sim$ 46$\%$) lie on or above the ``obscuration threshold", 
 while 
only $\sim$ 9 $\%$ of the moderate quasar DLAs lie on or above that threshold. About 77 $\%$ of the quasar super-DLAs lie above or within 1$\sigma$ of the obscuration 
threshold (including 2 of the 3 super-DLAs studied here), while only  $\sim$ 15 $\%$ of the moderate quasar DLAs do so. 
For GRB absorbers, the super-DLA and DLA fractions above or within 1$\sigma$ of  the threshold are more comparable, $\sim 75 \%$ and $60 \%$, 
respectively.  (We note that the 
fractions of GRB DLAs and super-DLAs above or within 1 $\sigma$ of the threshold would be larger if their $N_{\rm H I}$ 
values are larger by a factor of $\sim 2$.)    
(2) In this sense, the quasar super-DLAs are more similar to the GRB absorbers than are the 
moderate quasar DLAs. Indeed, given that the GRB DLAs are believed to arise in star-forming regions  within the main body of the galaxies (e.g., Pontzen et al. 2010), quasar super-DLAs are also likely to 
arise in quasar sightlines with small impact parameters to the absorber galaxies. (3) The H$_{2}$ covering fraction is significant for several super-DLAs above the 
``obscuration threshold", suggesting that they have higher molecular content than the moderate DLA population. This is consistent with the observations 
of H$_{2}$ in 5 out of 7 quasar super-DLAs by Noterdaeme et al. (2015b).  
Overall, it appears that the quasar super-DLAs are likely to arise more often in molecule-rich environments than the quasar moderate DLAs.  (See section 4.7 below for discussion of a search for molecular gas in 
the super-DLAs studied here.)

Fig. 11 shows the abundance of Fe relative to X, plotted as a function of the metallicity [X/H] (where X is taken to be either Zn, or S if Zn is not available), 
for quasar DLAs and super-DLAs. We have omitted the super-DLA toward Q2140-0321, because it has very large uncertainties ($> 0.9$ dex) in the 
[Zn/H] and [Fe/Zn] values, and no [S/H], [Fe/S] measurements are available.  Fig. 12 shows the abundance of Si relative to X, plotted as a function of the metallicity [X/H], for the DLAs and super-DLAs with Si and X detections 
(where, once again, X=Zn or S). 
Table 13 summarizes the statistics for the [Fe/X] vs. [X/H] data and the [Si/X] vs. [X/H] data for the moderate DLAs and super-DLAs. 

We first carried out bisector fits to the data in Figs. 11, 12. 
For the moderate DLAs, the intercepts for [Fe/X] and [Si/X] differ from each other at the $\sim$ 5.0 $\sigma$ level. This difference probably arises 
from the stronger depletion of Fe
than that of Si. The difference between the intercepts for the [Fe/X] and [Si/X] trends is less significant   ($\sim$ 2.7 $\sigma$) for super-DLAs; 
larger super-DLA samples are needed to determine whether this is an intrinsic difference or arises from the small sizes of the super-DLA samples.  About  19 $\%$ of the moderate DLA population and about 25$\%$ of the super-DLA population have [Si/X] $> 0$. This shows that 
the relative abundance patterns could be a combination of dust extinction and $\alpha$-enhancement. 
We also note that the bisector fit slope and intercept for the super-DLA trends are consistent with those for the 
moderate DLAs, to within the uncertainties allowed by the present sample sizes. Again, larger super-DLA samples are needed for more definitive determinations of 
the metallicity vs. depletion trends for super-DLAs. 

Next, we examined how the metallicity and depletion correlate for the moderate DLAs 
and super-DLAs, using the Spearman rank-order correlation test. The results of this non-parametric test are also summarized in Table 13. An anti-correlation between [Fe/X] 
and [X/H] has been noted in 
previous works (e.g., Meiring et al. 2006, 2009; Noterdaeme et al. 2008) and is indicative of 
increasing dust depletion with increasing metallicity. We confirm a strong anti-correlation for the moderate DLA population. For the smaller super-DLA sample, no correlation 
is seen. 
A [Si/X] vs. [X/H] anti-correlation is also seen at a significant level for the moderate DLAs.  A [Si/X] vs. [X/H] may be present  for the super-DLAs, but a larger super-DLA sample is needed to 
definitively determine whether such a correlation is indeed present, or whether it results from the small 
size of the current sample. It will also be interesting to determine [Fe/X] and [Si/X] in individual velocity components 
with higher resolution data to examine whether differences in the depletion pattern (such as those expected between cold and warm gas) are seen. 

\subsection{Search for Lyman-$\alpha$ Emission}

 The super-DLA toward Q1135-0010 shows strong emission lines of Ly-$\alpha$, H-$\alpha$, and [O~III], implying SFR $\sim 25$ M$_{\odot}$ yr$^{-1}$ 
 (Kulkarni et al. 2012; Noterdaeme et al. 2012b). This SFR level is much higher than that seen typically in the moderate DLA population. 
  A higher SFR could also indicate  
 a larger galaxy. The Ly-$\alpha$ 
 emission in the Q1135-0010 super-DLA shows symmetric double peaks, possibly arising in a starburst-driven outflow. In fact, stacked Super-DLA spectra show a
statistical detection of Ly-$\alpha$ emission at the bottom of the DLA trough, suggesting close association with Ly-$\alpha$ emitters (Noterdaeme et al. 2014). 
Such quasar super-DLAs resemble
the DLAs arising in GRB hosts (e.g., Guimaraes et al. 2012).

There is a hint of weak Ly-$\alpha $ emission near the center of the DLA trough for the systems toward Q0230-0334 and Q1418+0718. For Q0230-0334, 
on resampling the ESI spectrum to 1.5 {\AA} dispersion, we measure an integrated Ly-$\alpha$ emission flux of $(3.41 \pm 1.03) \times 10^{-17}$ erg s$^{-1}$ 
cm$^{-2}$. The SDSS spectrum of the same source gives an integrated Ly-$\alpha$ emission flux of $(3.66 \pm 1.50) \times 10^{-17}$ erg s$^{-1}$ cm$^{-2}$. 
For Q1418+0718, on resampling the UVES spectrum to 1.5 {\AA} dispersion, we measure an integrated Ly-$\alpha$ emission flux of 
$(1.81 \pm 0.80) \times 10^{-17}$ erg s$^{-1}$ cm$^{-2}$. 
These Ly-$\alpha$ fluxes would correspond to SFRs of $1.55 \pm 0.47$ M$_{\odot}$ yr$^{-1}$ based on the ESI data (or $1.66 \pm 0.66$ M$_{\odot}$ 
yr$^{-1}$ based on the SDSS data) for the absorber toward Q0230-0334,  assuming $L_{Ly-\alpha}/L_{H-\alpha}$= 8.7 for  case-B recombination, and adopting the Kennicutt (1998) relation 
between the SFR and H-$\alpha$ luminosity. The corresponding SFR estimate for the super-DLA toward Q1418+0718 is 
$0.74 \pm 0.33$ M$_{\odot}$ yr$^{-1}$. We note that these estimates assume no correction for dust absorption or resonant scattering by neutral gas.  
The SFRs would be higher if dust extinction is present. 

\subsection{Constraints on Electron Densities}

Fine structure lines, e.g. those of C II* or Si II*, allow constraints on the electron density, assuming equilibrium between collisional excitation
and spontaneous radiative de-excitation (and assuming collisional excitation rate for electrons to dominate over that of H atoms and over other excitation sources such as UV pumping 
and pumping from the cosmic microwave background) . While Si II* absorption has been detected in GRB DLAs, it is not common in quasar DLAs. 
Kulkarni et al. (2012) made the first detection of Si II$^{*}$ absorption in an intervening quasar DLA (in the super-DLA toward Q1135-0010). Unfortunately, no Si II$^{*}$ absorption is detected in any of 
the super-DLAs studied here. However, for the absorbers toward Q0234-0334 and Q0743+1421, we are able to constrain the electron density $n_{e}$ using the upper limits on Si II* and 
the measurements or lower limits on Si II. We use the Si II collisional
excitation rate $C_{12} = 3.32 \times 10^{-7} (T/10^{4})^{-0.5} \, {\rm exp}(-413.4/T)$ cm$^{3}$ s$^{-1}$
(see, e.g., Srianand \& Petitjean 2000) and the Si II* spontaneous radiative de-excitation rate 
$A_{21} = 2.13 \times 10^{-4} $ s$^{-1}$. 
For the super-DLAs toward 
Q0234-0334 and Q0743+1421, we obtain $n_{e} \le 0.16$ cm$^{-3}$ and $n_{e} \le 0.28$ cm$^{-3}$ , respectively, for an assumed temperature $T=500$ K,  or $n_{e} \le 0.28$ cm$^{-3}$ and 
$n_{e} \le 0.48$ cm$^{-3}$, 
respectively, for $T=7000$ K. For the absorber toward Q0743+1421, no constraint on C II*/C II could be obtained due to the 
saturation of C II* $\lambda 1336$ and C II $\lambda$ 1334 . For the absorber toward Q0234-0334, C II $\lambda 1334$ and C II* $\lambda 1336$ could not be 
measured due to severe blends with the Ly-$\alpha$ forest. 

The spectrum of Q1418+0718 shows a strong detection of C II$^{*}$ absorption, implying log  $N_{\rm C II^{*}}$ = 13.65.  The C II $\lambda$ 1334 line in this absorber is saturated, and a well-constrained estimate 
of $N_{\rm C II}$ could not be obtained from Voigt profile fitting. We therefore adopt the lower limit on $N_{\rm C II}$ implied by the apparent optical depth method 
(log $N_{\rm C II} \ge 14.84$). Combining these C II and C II$^{*}$ measurements, we estimate the electron density to be $n_{e} \le 0.51$ cm$^{-3}$ for an assumed temperature 
$T=500$ K or $n_{e} \le 1.03$ cm$^{-3}$ for $T=7000$ K. However, Si II* is not detected in this absorber. Based on the upper limit on Si II* and the detection of Si II, we obtain 
$n_{e} \le 0.035$ cm$^{-3}$  for $T=500$ K or $n_{e} \le 0.061$ cm$^{-3}$ for $T=7000$ K. 

Higher S/N and higher resolution spectra are essential to obtain more definitive constraints on the electron densities in our super-DLAs.  Shorter-wavelength 
observations covering C II $\lambda$ 1036.3 and C II* $\lambda$ 1037.0 would also be useful to obtain additional constraints on the C II and C II* column densities. 
We note, nevertheless, a couple of points. First, differences between $n_{e}$ derived from different elements have been observed for the Galactic diffuse ISM (e.g., Welty et al. 2003). 
Welty et al. found a mean $n_{e} \sim 0.14 \pm 0.07$ cm$^{-3}$ for C, Na, K. Furthermore, they found that $n_{e}$ derived from the Ca~I/ Ca~II ratio is often much greater than the value expected 
if most of the electrons come from the photo-ionization of C, and suggested that the ionization balance of heavy elements in the diffuse ISM is affected by additional processes besides photoionization and 
radiative recombination. Second, we also note that the median $n_{e}$ in DLAs has been found to be $0.0044 \pm 0.0028$ cm$^{-3}$ in a recent study (Neeleman et al. 2015). 
This is much lower than the electron density found for the super-DLA toward Q1135-0010 ($n_{e} = 0.53-0.91$ cm$^{-3}$; Kulkarni et al.  2012). 
It would be interesting to 
investigate with higher S/N, higher-resolution data targeting fine-structure lines for a larger number of super-DLAs whether super-DLAs in general 
have higher electron densities than moderate DLAs. A higher electron density may indicate a higher degree of ionization, possibly arising from a higher degree of star formation compared to the moderate 
DLAs. 

\subsection{Cooling Rate and Star Formation Rate Density}

The C II$^{*}$ column densities can also be used to constrain the cooling rates and the star formation rate (SFR) densities (Wolfe et al. 2003).  For the super-DLA toward Q1418+0718, 
using the measurement of $N_{\rm C II*}$, we estimate the cooling rate $l_{c} = N_{\rm C II*} h \nu_{ul} A_{ul}/ N_{\rm H I}$, where $\nu_{ul}$ and $A_{ul}$ 
denote the frequency and the coefficient for spontaneous photon decay for the ${2}P_{3/2}$ to $^{2}P_{1/2}$ transition (e.g., Pottasch et al. 1979). Using 
$A_{ul} = 2.29 \times 10^{-6}$, we estimate $l_{c} = 2.57 \times 10^{-28}$ erg s$^{-1}$ per H atom. This is substantially lower 
than the cooling rate in the super-DLA toward Q1135-0010 (which was estimated by Kulkarni et al. 2012 to be in the range $2.6 \times 10^{-27} < l_{c} < 1.2 \times 10^{-25}$ erg s$^{-1}$ per H atom) , and 
suggests a lower SFR in the super-DLA toward Q1418+0718. 

To estimate the SFR density, we use the fact that the Fe abundance, the Si abundance, and the cooling rate for 
the super-DLA toward Q1418+0718 are within 0.05-0.26 dex of those for the $z=2.309$ DLA toward Q0100+13 from Wolfe et al. (2003). For the latter system, Wolfe et al. 
estimate log  $\dot{\psi_{*}} \sim -2.8$ M$_{\odot}$ yr$^{-1}$ kpc$^{-2}$ for the cold neutral medium (CNM) and log  $\dot{\psi_{*}} \sim -1.6$ M$_{\odot}$ yr$^{-1}$ kpc$^{-2}$ for the warm neutral medium 
(WNM). We therefore estimate the SFR density in the 
super-DLA toward Q1418+0718 to be approximately  log  $\dot{\psi_{*}}  \sim -3$ M$_{\odot}$ yr$^{-1}$ kpc$^{-2}$ for CNM or log  $\dot{\psi_{*}}  \sim -1.9$ M$_{\odot}$ yr$^{-1}$ kpc$^{-2}$ for WNM, 
roughly an order of 
magnitude lower than the SFR density range 
estimated for the super-DLA toward Q1135-0010 (Kulkarni et al. 2012). For comparison, we note that the Kennicutt (1998) relation 
$\dot{\psi_{*}} = [2.5 \times 10^{-4}] \times [N_{\rm H I} / 1.26 \times 10^{20}]^{1.4}$ would imply log  $\dot{\psi_{*}}  \sim -1.4$ M$_{\odot}$ yr$^{-1}$ kpc$^{-2}$ for the super-DLA toward Q1418+0718. 
Measurements of  C II* absorption 
in a larger sample of super-DLAs are essential to systematically compare their SFR densities with those in other DLAs and in near-by galaxies. 

An SFR density of log  $\dot{\psi_{*}}  \sim -2.5$ M$_{\odot}$ yr$^{-1}$ kpc$^{-2}$, together with the SFR = $0.74$ M$_{\odot}$ yr$^{-1}$ estimated above from the marginal Ly-$\alpha$ emission, 
 would imply the line-of-sight size of the absorber toward Q1418+0718 to be $\sim 15$ kpc, if the star formation were spread uniformly within the absorber. This is much larger than the sizes observed 
 ($\sim$few kpc) for star-forming galaxies at high redshift. However, the absorber could be considerably smaller in size depending on 
 the clumpiness of the star formation and the relative WNM content.  It would be very interesting to obtain integral field spectroscopic observations of super-DLAs, similar to those obtained 
 for lower column density DLAs and sub-DLAs by P\'eroux et al. (2011, 2012, 2013) to study the distribution of star formation and metallicity in these objects. 
 
\subsection{Search for Molecules}

For Q0230-0334 and Q0743+1421, the wavelength coverage was not adequate to examine the Lyman or Werner band absorption lines of H$_{2}$. For Q1418+0718, the wavelength coverage 
includes H$_{2}$ Lyman band transitions in the range 1109-1152 {\AA} in the absorber rest frame, but the S/N in this region is very low. 
No lines for the $H_{2}$ J1-J7 rotational levels were detected; the non-detections of the strongest 
transitions covered imply 3 $\sigma$ upper limits of  log $N_{\rm H2 J1} < 15.55$, 
 log $N_{\rm H2 J2} < 15.66$,  log $N_{\rm H2 J3} < 15.77$,  log $N_{\rm H2 J4} < 15.84$, log $N_{\rm H2 J5} < 15.16$, 
log $N_{\rm H2 J6} < 15.23$, and log $N_{\rm H2 J7} < 15.02$, respectively. The total of the H$_{2}$ J1 to J7 column densities 
log $N_{\rm H2 J1-J7} < 16.40$ 
may suggest a low molecular content. 

For all three of our absorbers, several 
lines of CO were covered, but not detected. From the non-detections of the strongest lines covered (near 1477 {\AA} in the absorber rest frame), we estimate 3 $\sigma$ upper limits of log $N_{\rm CO J0} < 13.41$, log $N_{\rm CO J1} < 13.71$, 
and log $N_{\rm CO J2} < 13.71$ for the super-DLA toward Q0230-0334. For the absorber toward Q0743+1421, from the non-detection of the lines near 1477 and 1544 {\AA}, we estimate 3 $\sigma$ upper limits of log $N_{\rm CO J0} < 13.56$, log $N_{\rm CO J1} < 13.86$, 
log $N_{\rm CO J2} < 13.86$, log $N_{\rm CO J3} < 14.12$, log $N_{\rm CO J4} < 14.12$, log $N_{\rm CO J5} < 14.12$, and log $N_{\rm CO J6} < 14.12$, and a total log $N_{\rm CO J0-J6} < 14.85$. 
For the absorber toward Q1418+0718, we estimate 3$\sigma$ upper limits of log $N_{\rm CO J0} < 13.68$, log $N_{\rm CO J1} < 14.02$, log $N_{\rm CO J2} < 13.98$, log $N_{\rm CO J3} < 14.63$, log $N_{\rm CO J4} < 14.63$, 
log $N_{\rm CO J5} < 14.64$, and log $N_{\rm CO J6} < 14.65$, from the non-detections of the strongest transitions covered (near 1368, 1419, and 1447 {\AA}).  

Noterdaeme et al. (2015b) reported strong H$_{2}$ detections with log $N_{H_{2}} \sim 17.1-20.1$ in 5 out of 7 
super-DLAs, and log $N_{H_{2}} < 14.6$ in the remaining two super-DLAs. Higher S/N and lower wavelength spectra of our super-DLAs are needed to obtain more definitive determinations 
of their H$_{2}$ and CO contents. If our super-DLAs indeed have low molecular contents despite having log $N_{\rm H I} > 21.7$, that would contrast with observations of the Galactic ISM, which shows a sharp increase in H$_{2}$ column densities at 
log $N_{H I} > 20.7$ (e.g., Savage et al. 1977); however, low molecular contents would be consistent with the low SFRs in our super-DLAs suggested by 
the marginal or weak detections of Ly-$\alpha$ emission in the DLA troughs. 

\subsection{Gas Kinematics}

A correlation between metallicity and gas velocity dispersion has been noted for DLAs (e.g., Ledoux et al. 2006; Meiring et al. 2007; Som et al. 2015). Such a 
correlation could indicate a mass-metallicity relation for the absorbing galaxies, if the gas velocity dispersion is a measure of the mass of the absorber. Alternately, 
the velocity dispersion could reflect turbulent motions or outflows. 

Fig. 13 shows our measurements of the velocity dispersion $\Delta v_{90}$ for each of the three super-DLAs studied here. 
These measurements were made on well-detected, unsaturated lines such as Cr II $\lambda$2056 or Fe II $\lambda$2250. The velocity dispersions were found to be 
$\Delta v_{90}$ = 149.9 km s$^{-1}$, 67.9  km s$^{-1}$ and 31.8 km s$^{-1}$, respectively, for the absorbers toward Q0230-0334, Q0743+1421, and Q1418+0718.  (While higher resolution spectra would give more reliable determinations of $\Delta v_{90}$ for the 
absorbers toward Q0230-0334 and Q0743+1421, we regard our values for these systems as reasonable approximations, given that they are quite distinct from each other and 
feasible at the spectral resolution of our data.) It is surprising that two of the 
three systems show such low velocity dispersions. A low velocity dispersion $\Delta v_{90}$ = 28 km s$^{-1}$ was also found for the $z_{abs} = 2.34$ super-DLA with log $N_{\rm H I} = 22.4$ 
toward Q2140-0321 (Noterdaeme et al. 2015a). 

To compare the kinematics of regular DLAs and super-DLAs, we plot in Fig. 14 the velocity dispersion vs. the H I column density. To our knowledge, the velocity dispersion has been 
published for only six other super-DLAs with $z_{abs} < z_{em}$ besides the three from this study. For one more super-DLA, we estimated $\Delta v_{90}$ approximately from the velocity structure provided in Guimaraes et al. (2012). The vertical 
and horizontal dashed lines in Fig. 14 denote log $N_{\rm H I} = 21.7$ and $\Delta v_{90} = 160$ km s$^{-1}$. It is interesting to note that  20\% of the moderate DLAs (those with log 
$N_{\rm H I} < 21.7$) have 
$\Delta v_{90} >160$ km s$^{-1}$, while 10$\%$ of the super-DLAs have such large velocity dispersions.  While  52$\%$ of the DLAs with log $N_{\rm H I}$$<$21.7 have 
$\Delta v_{90} >90$ km s$^{-1}$,  40$\%$ of the super-DLAs have $\Delta v_{90} >90$ km s$^{-1}$. Since the sample of super-DLAs with $\Delta v_{90}$ measurements is so small 
(10  systems), we also compare the samples with log $N_{\rm H I} < 21.5$ and log $N_{\rm H I} \ge 21.5$. The mean velocity dispersion is 125.6 km s$^{-1}$ for systems 
with log $N_{\rm H I} < 21.5$, and 79.2  km s$^{-1}$ for systems with log $N_{\rm H I} \ge 21.5$. A two-sample Kolmogorov-Smirnov test between DLAs with log $N_{\rm H I} < 21.5$ and those with 
log $N_{\rm H I} \ge 21.5$ gives $D =$ 0.309 ($D$ being the K-S statistic, i.e. the maximum absolute value of the difference between the cumulative distribution functions of $\Delta v_{90}$ for 
the two samples), with a probability of  0.815 that the two samples are not similar. 

Fig. 15 shows the metallicity vs. log of the velocity dispersion for moderate DLAs and super-DLAs.  Once again, we exclude the super-DLA  toward Q2140-0321, because it has a very large 
uncertainty ($> 0.9$ dex) in [Zn/H]  and no [S/H] measurement. The data for the moderate DLAs indeed show a  significant correlation with a Spearman rank-order correlation 
coefficient $r_{S}$ = 0.653 and a probability $P$ $<$ $1\times 10^{-6}$ of no correlation. The super-DLA data do not 
show a strong correlation ($r_{S}$ = 0.371, $P$= 1.46$ \times 10^{-1}$), which could be a result of the fact that they occupy a narrow range in metallicity compared to moderate DLAs.  The fact that most 
super-DLAs have metallicities below -1.0 dex may also explain the small fraction of super-DLAs with high velocity dispersions. Measurements for more quasar super-DLAs 
are essential to determine whether there is a statistically significant difference between the velocity dispersion vs. metallicity relations for quasar super-DLAs and other (moderate) quasar DLAs. If quasar super-DLAs indeed have 
smaller velocity dispersions on average than other quasar DLAs, this may argue against the possibility that the quasar super-DLAs arise in larger, more massive galaxies, if the velocity dispersion is a measure of the mass.  Alternately, it may suggest the presence of cooler or less turbulent 
gas in quasar super-DLAs than in other quasar DLAs. 
In this context, it is interesting to note that Arabsalmani et al. (2015) have recently examined the velocity dispersion vs. metallicity relation for long-duration GRB host galaxies, 
and found it to be very similar to that for quasar DLAs. Moreover, they find a weak evidence for the velocity dispersions being larger in 
GRB DLAs. It is therefore important to expand the velocity dispersion and metallicity samples for quasar super-DLAs in order to understand how they compare to GRB DLAs.

\section{Summary}

We have analyzed Keck ESI and VLT UVES spectra of three super-DLA absorbers and measured abundances of a number of elements. All three absorbers show 
remarkably similar metallicities of $\sim -1.3$ to $\sim -1.5$ dex and comparable, definitive depletion levels, as judged from [Fe/Zn] and [Ni/Zn].  Two of the absorbers show 
supersolar [S/Zn] and [Si/Zn]. Combining our measurements with 
those for other super-DLAs and moderate DLAs from the literature, we find that the correlation between Fe  (as well as Si) depletion and metallicity is weaker for the super-DLAs than for the 
moderate DLAs. The offset between the best-fitting [Si/X] vs. [X/H] and [Fe/X] vs. [X/H] trends indicates the greater depletion of Fe compared to that of Si. Using 
the fine structure lines of C II* and Si~II*, we have constrained the electron densities in the absorbers. While we can only put upper limits on the electron densities, these limits are higher than 
the median electron densities in the general DLA population; this, together with comparable $n_{e}$ values seen in a few other super-DLAs, could mean the presence of denser and/or more ionized gas in the super-DLAs.  
Using potential detections of weak Ly-$\alpha$ emission 
at the bottom of the DLA troughs, we estimate SFR in the absorbers toward Q0230-0334 and Q1418+0718 to be  $\sim$1.6 and $\sim$0.7 M$_{\odot}$ yr$^{-1}$, respectively. 
For the absorber toward Q1418+0718, the C II* measurement suggests a star formation rate density of log  $\dot{\psi_{*}}  \sim -3$ M$_{\odot}$ yr$^{-1}$ kpc$^{-2}$ for CNM or 
log  $\dot{\psi_{*}}  \sim -1.9$ M$_{\odot}$ yr$^{-1}$ kpc$^{-2}$ for WNM. Finally, measurements of the 
velocity spread $\Delta v_{90}$ suggest that super-DLAs may have narrower velocity dispersions and may arise in cooler and/or less turbulent gas. 

Our study has 
further demonstrated the potential of super-DLAs as unique laboratories to study the physical and chemical properties of potentially star-forming interstellar gas in distant galaxies. Observations of a 
larger super-DLA sample at higher S/N are essential to further understand the nature of these unique absorbers and how their underlying galaxies compare with those probed by the moderate  DLAs, 
and also those probed by the GRB DLAs.  

\acknowledgments
 We thank the referee for constructive comments that have helped to improve this paper. 
VPK, DS, and SM gratefully acknowledge support from the National Science Foundation grant AST/1108830 and NASA   grant  
NNX14AG74G (PI Kulkarni).  Additional support from NASA Space Telescope Science Institute grant HST-GO-12536.01-A and NASA Herschel Science Center grant 1427151 (PI Kulkarni) 
is gratefully acknowledged. DS also acknowledges support from the A*MIDEX project (ANR- 11-IDEX-0001-02) funded by the 
``Investissements d'Avenir" French Government program, 
managed by the French National Research Agency (ANR). CP thanks the European Southern Observatory science visitor program for support.  
The authors wish to recognize and acknowledge the very significant cultural role and reverence that the summit of Mauna Kea has always had within the indigenous Hawaiian community.  
We are most fortunate to have the opportunity to conduct observations from this mountain. 

{\it Facilities:} \facility{VLT (UVES)}, \facility{Keck (ESI)}.

\clearpage

\begin{figure}
\epsscale{.80}
\plotone{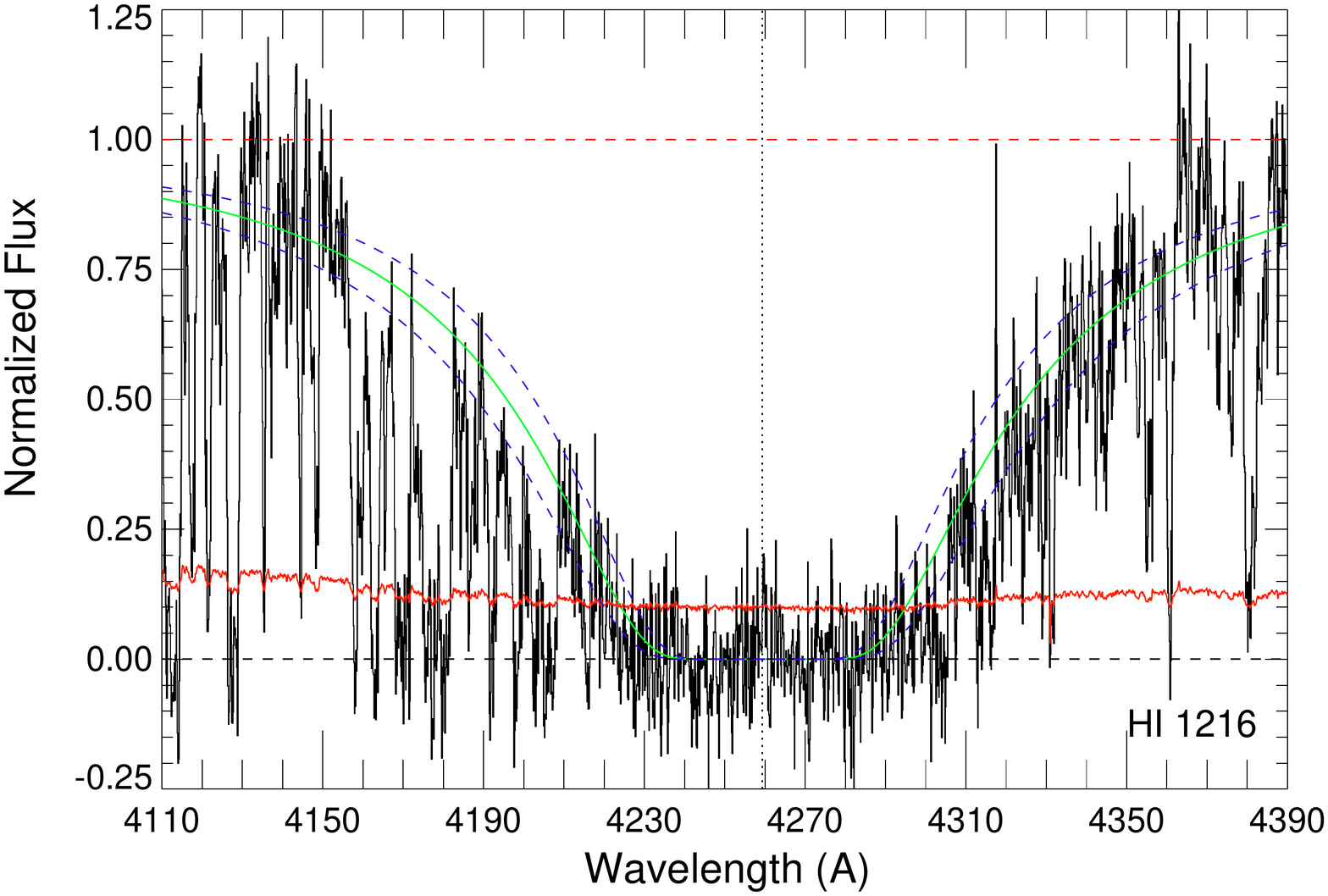}
\caption{H I Ly-$\alpha$ absorption feature in our ESI data for the $z=2.5036$ absorber toward J0230-0334. The continuum-normalized flux 
is shown in black, while the $1\sigma$ error array in the 
normalized flux is shown in red. The solid green curve represents the best-fitting Voigt profile corresponding to log $N_{\rm H I}$=21.74 and the
 dashed blue curves denote the
profiles corresponding to estimated $\pm 1\sigma$ deviations ($\pm 0.1$ dex) from the best-fitting $N_{\rm H I}$ value. The horizontal dashed red line denotes 
the continuum level. The vertical dotted black
line denotes the center of the Ly-$\alpha$ line. \label{fig1}}
\end{figure}

\begin{figure}
\epsscale{.80}
\plotone{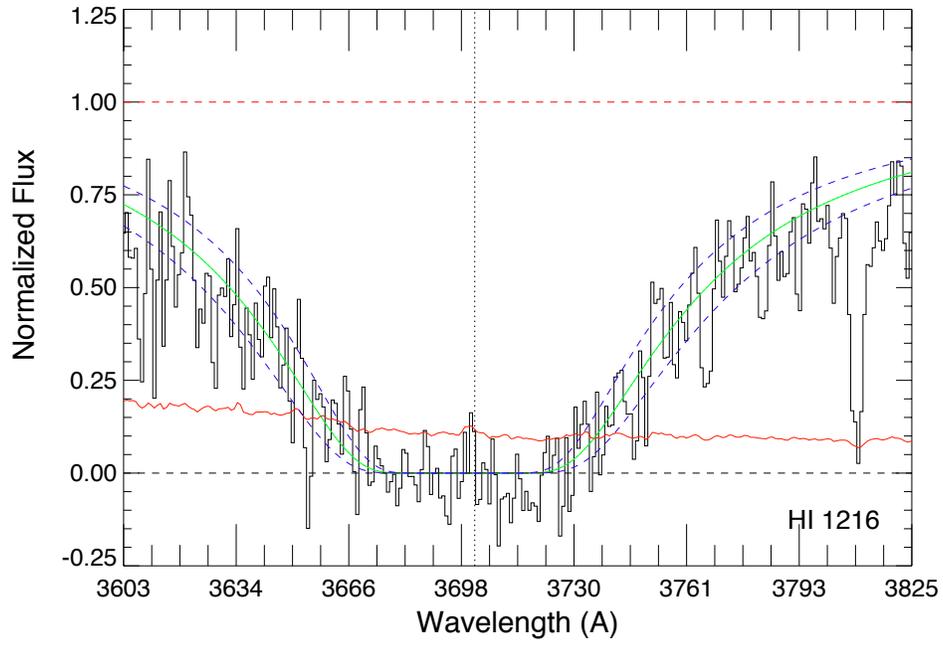}
\caption{Same as Fig. 1, but for the $z_{abs}=2.045$ super-DLA toward Q0743+1421.  The solid green and 
dashed blue profiles correspond to log $N_{\rm H I} = 21.9 \pm 0.1$. \label{fig2}}
\end{figure}

\begin{figure}
\includegraphics[angle=270,scale=.50]{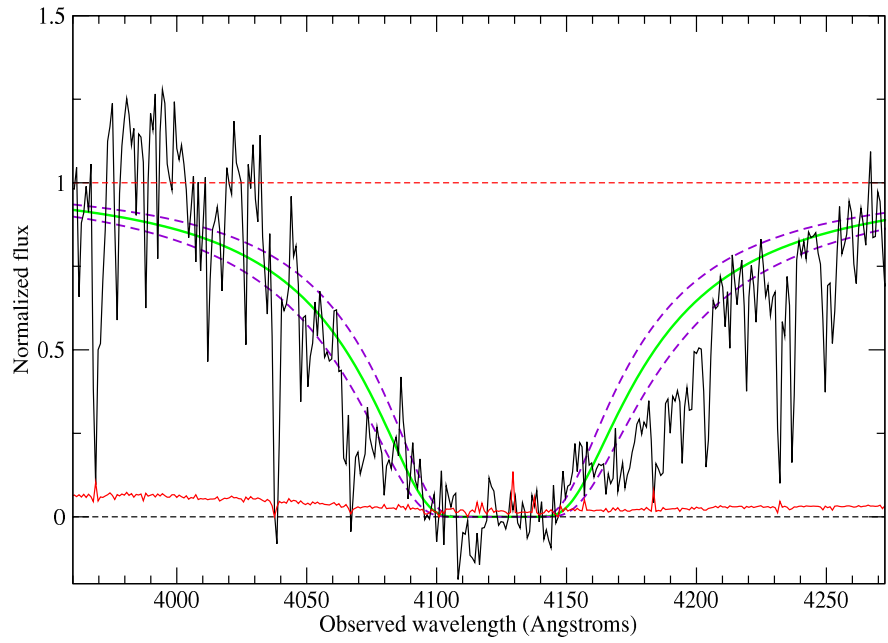}
\caption{Same as Fig.1, but for the $z=2.392$ super-DLA toward Q1418+0718.  The solid green and dashed 
blue profiles correspond to log $N_{\rm H I} = 21.7 \pm 0.1$. \label{fig3}}
\end{figure}

\begin{figure}
\epsscale{0.8}
\plotone{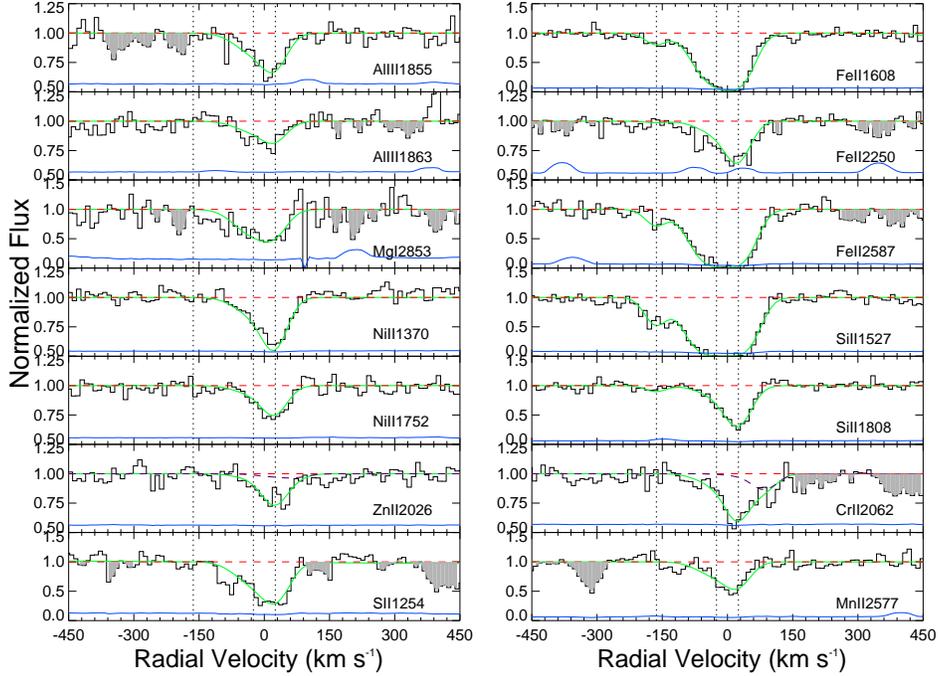}
\caption{Velocity plots for several metal lines of interest for the  $z=2.5036$ system in the spectrum of Q0230-0334. In each panel,  the normalized data 
are shown in black, the
solid green curve indicates the theoretical Voigt profile fit to the absorption feature, and the dashed red line shows the continuum level. 
The $1\sigma$ error values in the normalized flux
are represented by the blue curves near the bottom of each panel. Note that in a few panels with weak lines, if the normalized flux scale shown 
starts at 0.5, the 1-$\sigma$ error arrays have been offset by 0.5, so that they can be viewed in the same panels. The vertical dotted lines indicate the positions of the components that were 
used in the fit. In the `ZnII~2026' panel, the
solid green curve represents the combined contributions from Zn II $\lambda$ 2026.1 and Mg I $\lambda$ 2026.5 lines while the contribution 
from Mg~I $\lambda$ 2026.5 alone to this blend,
determined from the Mg I $\lambda$ 2853 line, is represented by the dashed purple curve. The solid green curve in the `CrII~2062' panel 
represents the combined contributions from
Cr~II $\lambda$ 2062.2 and Zn~II $\lambda$ 2062.7 lines and the contribution from the Zn~II $\lambda$ 2062.7 line alone is shown using 
a dashed purple line. The regions shaded in gray in some
of the panels represent absorption unrelated to the line presented or regions of high noise. 
 \label{fig4}}
\end{figure}

\begin{figure}
\epsscale{.80}
\plotone{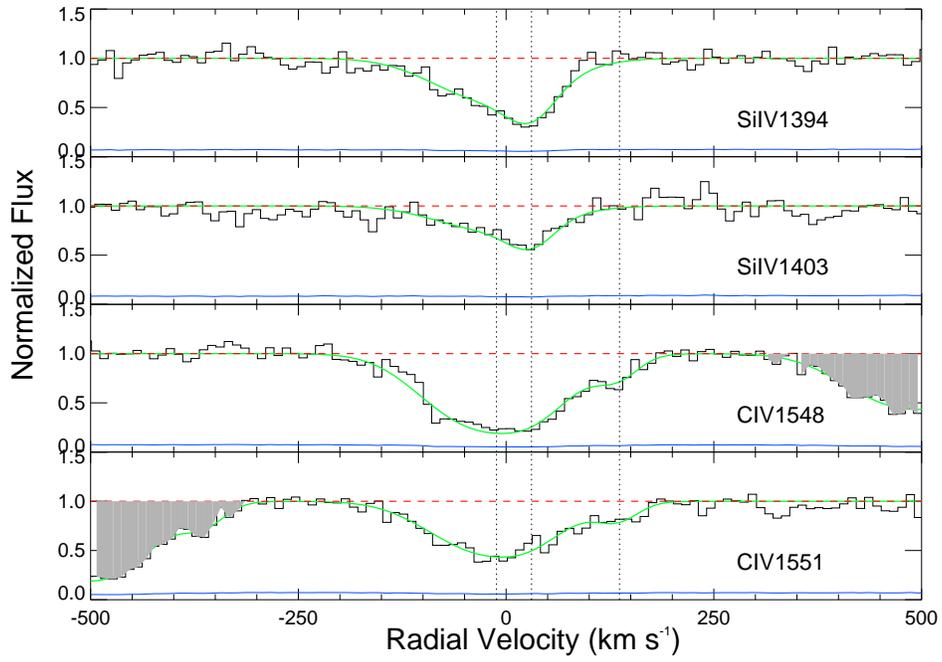}
\caption{Same as Fig. 4, but for the high-ionization metal absorption lines  in the $z=2.5036$ super-DLA toward Q0230-0334.   \label{fig5}}
\end{figure}

\begin{figure}
\epsscale{.80}
\plotone{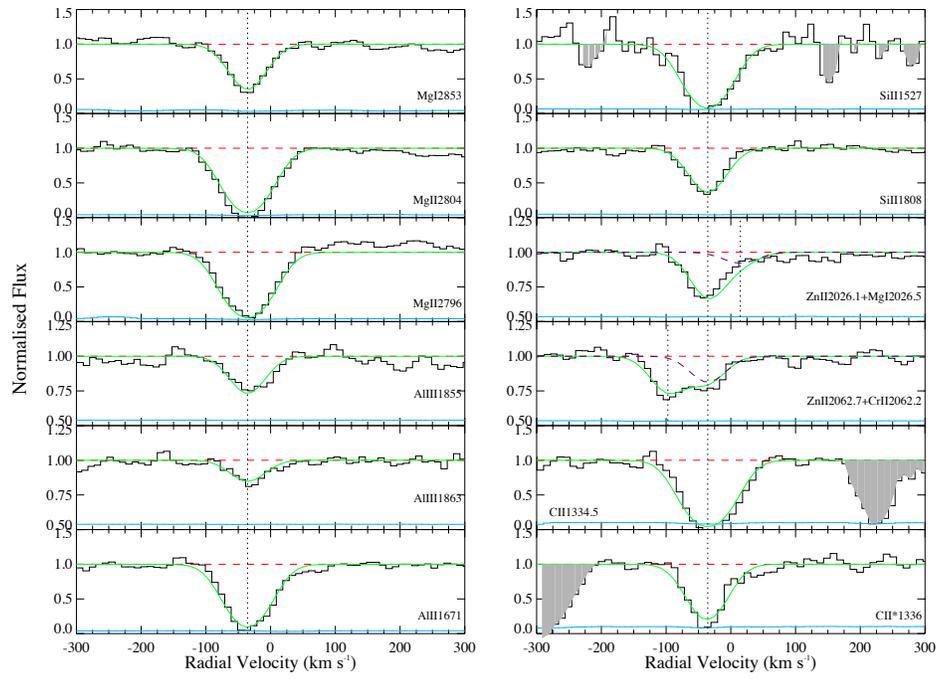}
\caption{Same as Fig.4, but for key low-ionization metal absorption lines in the $z_{abs}=2.045$ super-DLA toward Q0743+1421.  \label{fig6}}
\end{figure}

\begin{figure}
\epsscale{.80}
\plotone{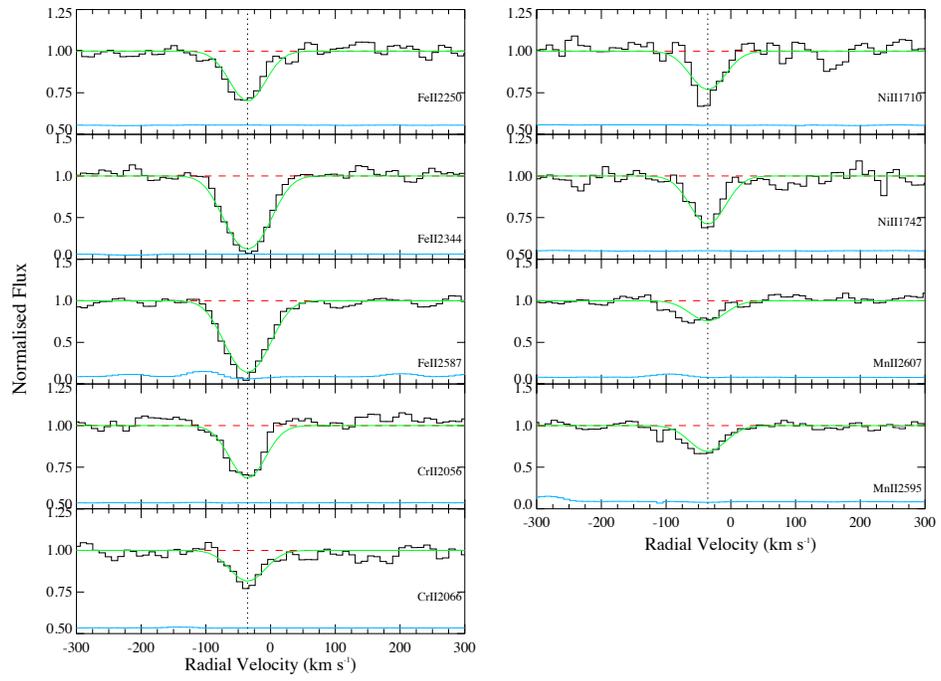}
\caption{Same as Fig. 4, but for additional low-ionization metal absorption lines in the $z_{abs}=2.045$ super-DLA toward Q0743+1421.  \label{fig7}}
\end{figure}

\begin{figure}
\epsscale{1.0}
\plotone{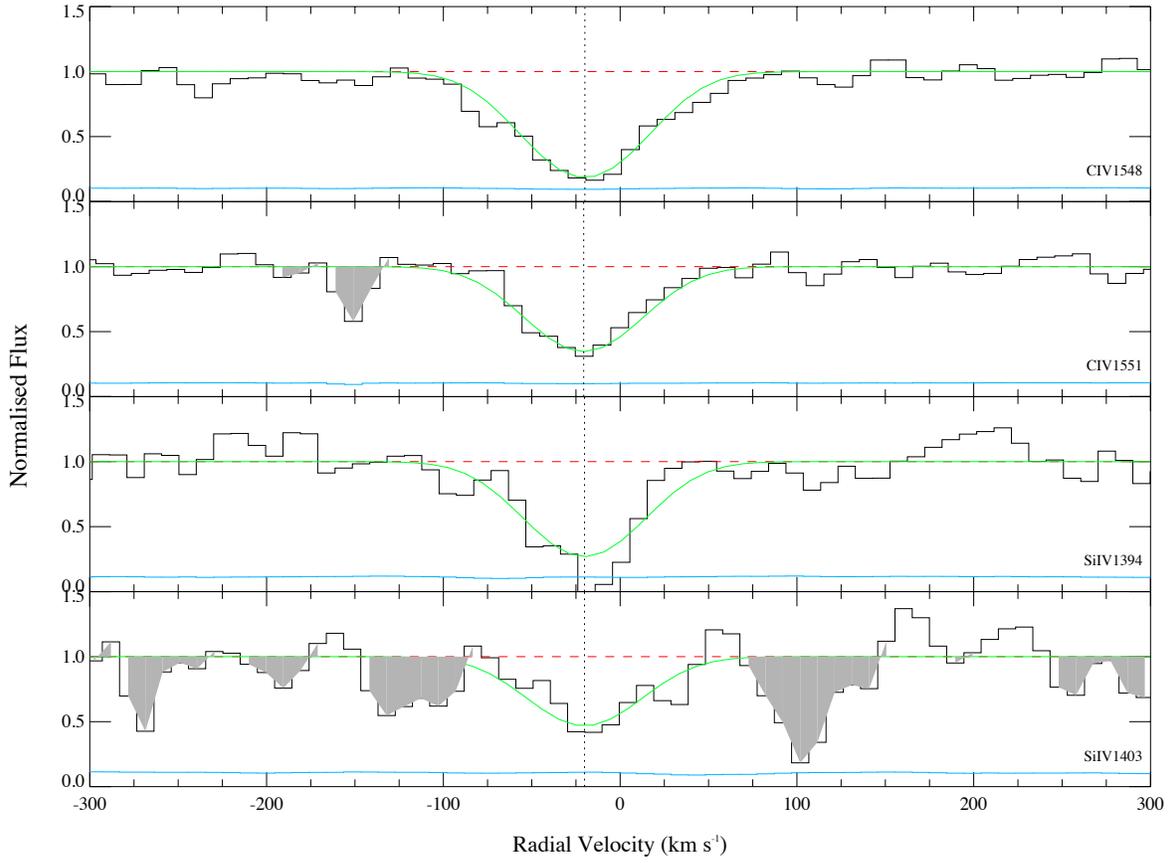}
\caption{Same as Fig. 4, but for the high-ionization lines  in the $z_{abs}=2.045$ super-DLA toward Q0743+1421.   \label{fig8}}
\end{figure}

\begin{figure}
\epsscale{1.04}
\plotone{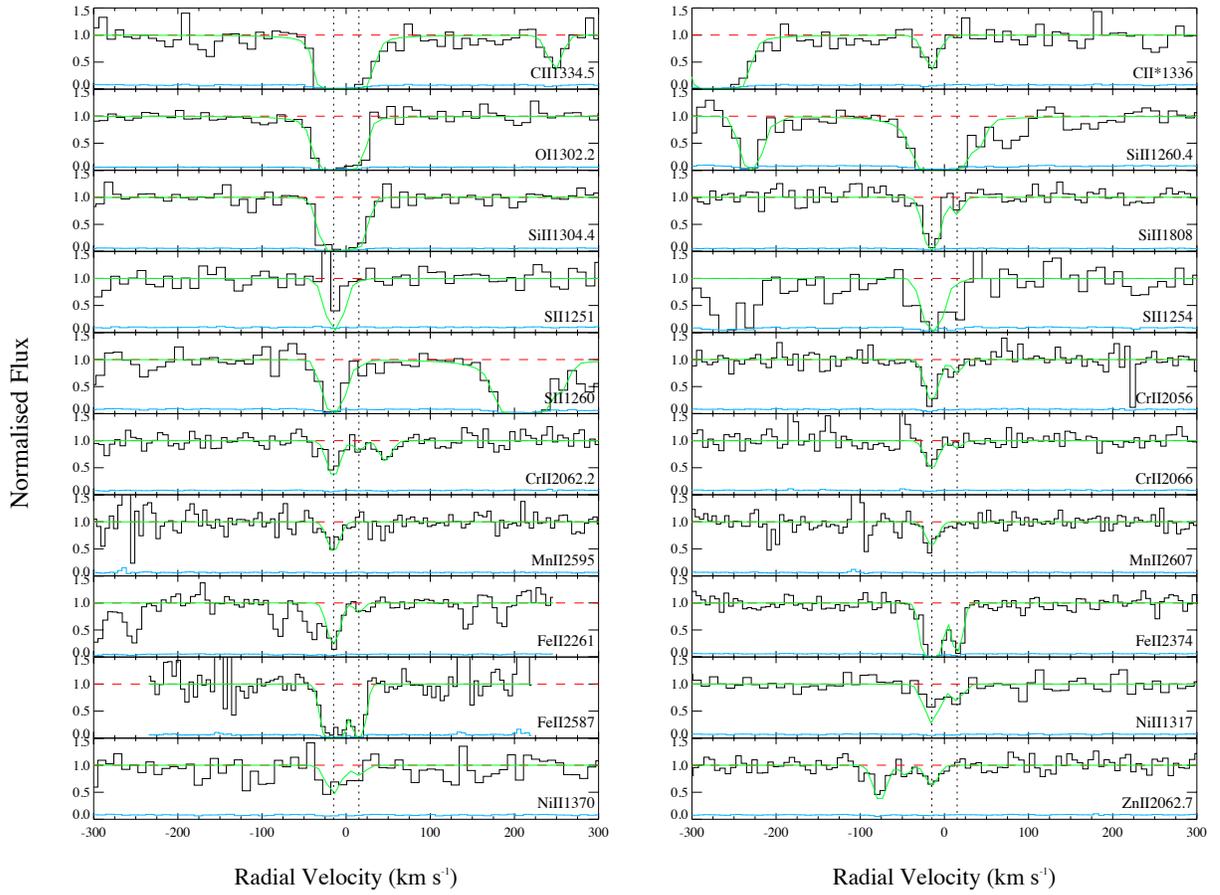}
\caption{Same as Fig.4, but for key metal absorption lines in the $z=2.392$ super-DLA toward Q1418+0718.  \label{fig9}}
\end{figure}

\clearpage

\begin{figure}
\epsscale{1.00}
\plotone{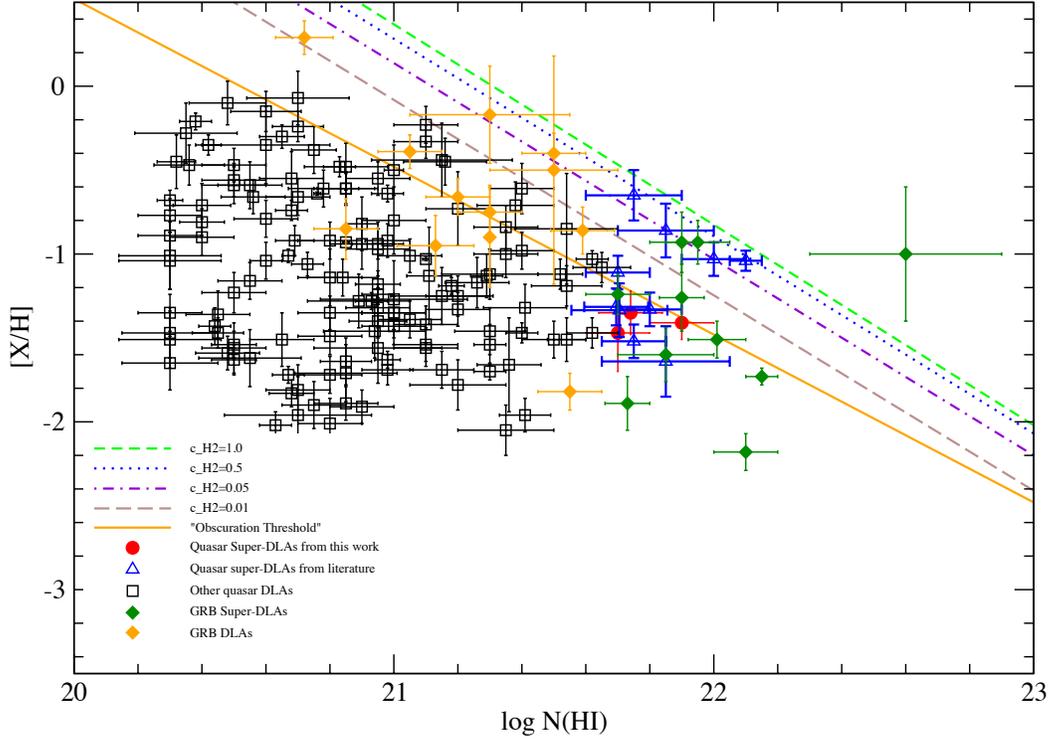}
\caption{Plot of metallicity (based on Zn or S absorption) vs. H  I column density for quasar super-DLAs from this study and 
the literature, other quasar DLAs, and GRB DLAs/ super-DLAs. Short-dashed, dotted, dot-dashed, and long-dashed curves show trends expected for molecular 
Hydrogen core ``covering fractions" of 1.0, 0.5, 0.05, and 0.01, adopted from Krumholz et al. (2009a). The solid orange 
line shows  the ``obscuration threshold'' of Boisse et al. (1998). 
 \label{fig10}}
\end{figure}

\begin{figure}
\epsscale{1.00}
\plotone{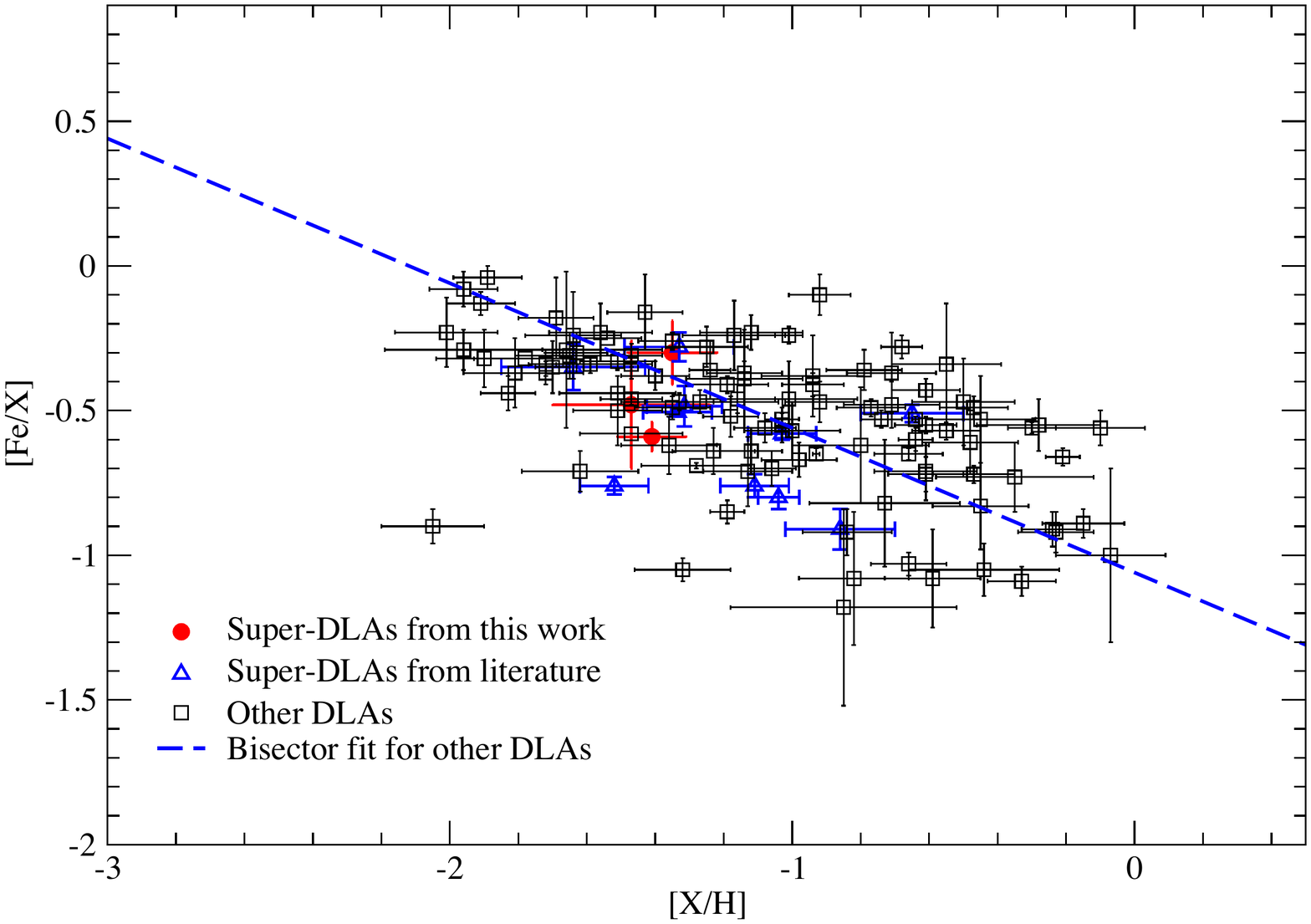}
\caption{Plot of the depletion of Fe relative to Zn or S vs. the metallicity based on Zn or S, for quasar super-DLAs from this study and the literature, and for 
other quasar DLAs.  The dashed line shows the bisector fit for the other quasar DLAs.
 \label{fig11}}
\end{figure}

\begin{figure}
\epsscale{1.00}
\plotone{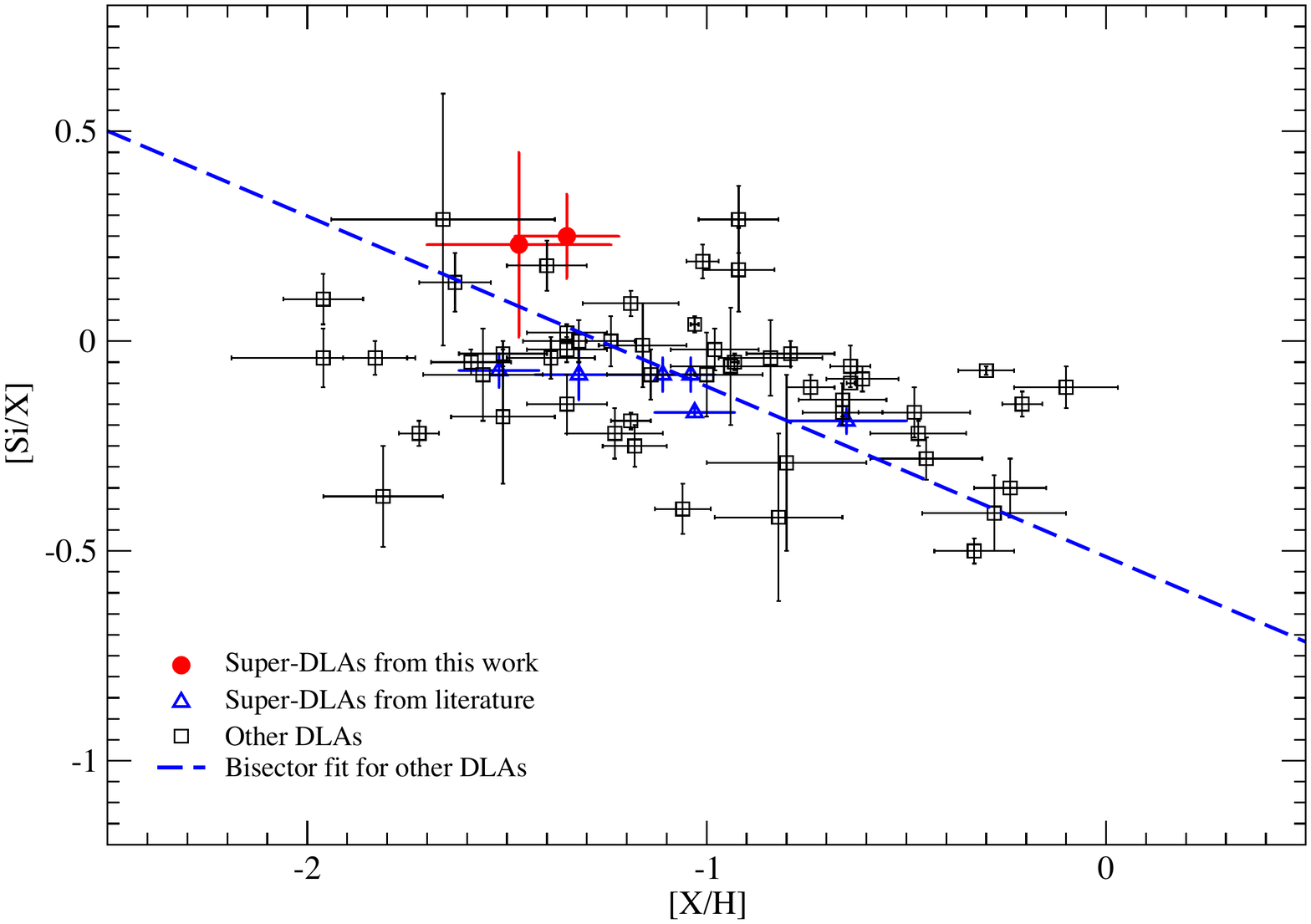}
\caption{Plot of the depletion of Si relative to Zn or S vs. the metallicity based on Zn or S, for quasar super-DLAs from this study and the literature 
 and other quasar DLAs.  The dashed line shows the bisector fit for the other quasar DLAs.
 \label{fig12}}
\end{figure}

\begin{figure}
\epsscale{1.00}
\plotone{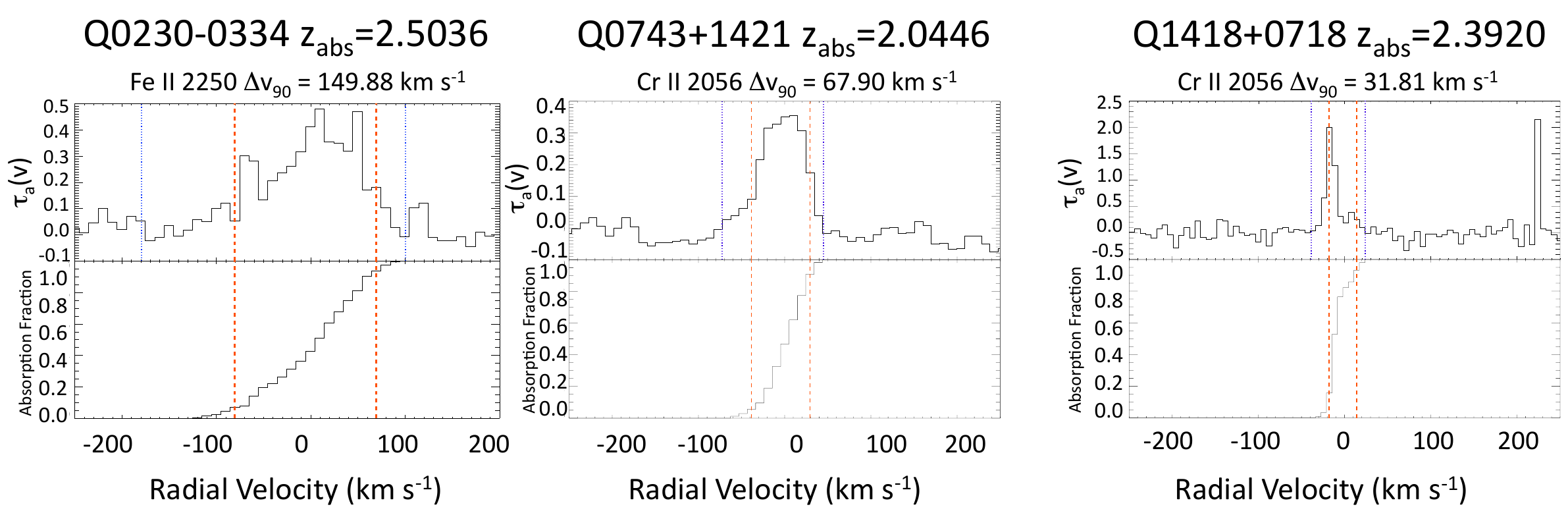}
\caption{Estimates of velocity dispersions for a representative line in each of the three super-DLAs observed in this study.  The top and bottom sub-panels 
for each absorption line shows the apparent optical depth profile and the absorption fraction as functions of the velocity of the 
absorbing gas relative to the stated redshift. 
The red dashed vertical lines mark the 5\% and 95\% levels of absorption. The blue dotted vertical lines in each of the top 
sub-panels show the 
full range of velocities used to make the measurements. 
 \label{fig13}}
\end{figure}

\begin{figure}
\epsscale{1.00}
\plotone{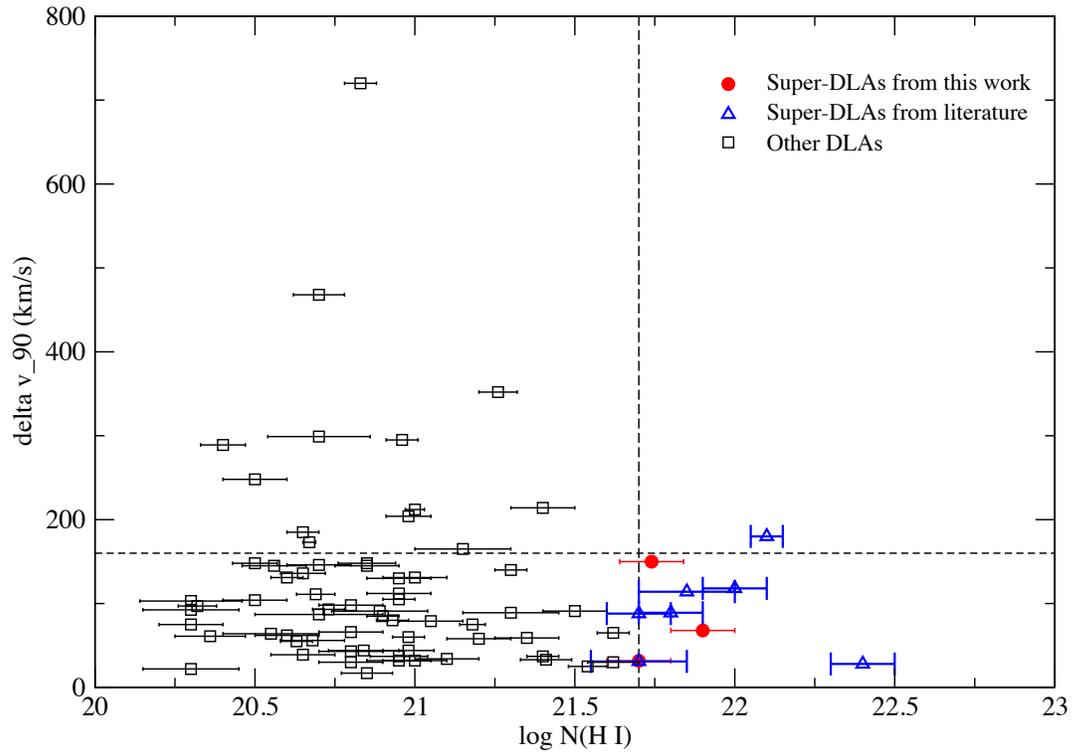}
\caption{Plot of velocity dispersion vs. H I column density for DLAs and super-DLAs. Super-DLAs appear to have a  somewhat smaller 
incidence of large velocity dispersions than other DLAs. 
 \label{fig14}}
\end{figure}

\begin{figure}
\epsscale{1.00}
\plotone{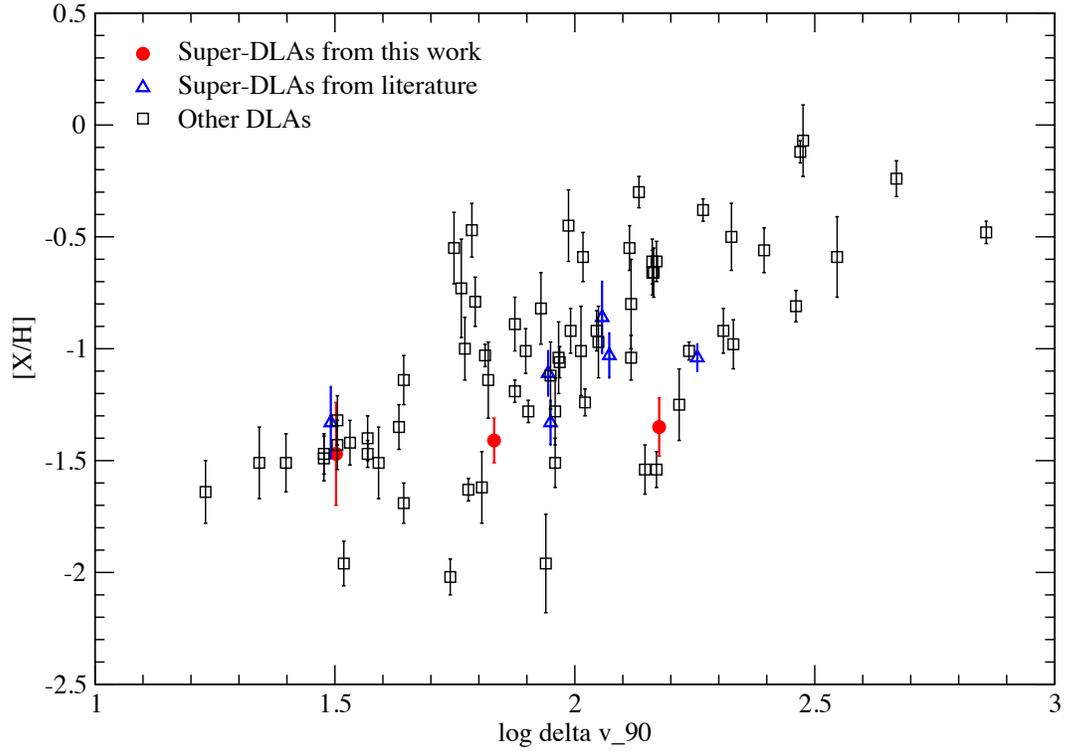}
\caption{Gas-phase metallicity vs. log of the velocity dispersion (in km s$^{-1}$) for DLAs and super-DLAs. While other DLAs show a correlation between 
these quantities, the small super-DLA sample does not show a strong correlation. 
 \label{fig15}}
\end{figure}

\clearpage

\begin{table}
\begin{center}
\caption{Targets Observed}
\vskip 10pt
\begin{tabular}{cccccc}
\tableline\tableline
Quasar	& RA,DEC(J2000)& $z_{em}$ & $z_{abs}$& Instrument & Integration time\\
\tableline
Q0230-0334&02:30:11.30-03:34:50.2&2.872&2.5036&Keck/ESI&$2100$ s $\times 1$\\
Q0743+1421&07:43:44.26+14:21:34.9&2.281&2.045&Keck/ESI&$1800$ s $ \times 2$ + 1300 s\\
Q1418+0718&14:18:01.86+07:18:43.6&2.572&2.3920&VLT/UVES&$5400$ s $\times 3$\\
\tableline
\end{tabular}
\end{center}
\end{table}
\clearpage

\begin{landscape}																																																																																																																																																																																																																																																																																																																																																																																																																																																																																																																																																																																																																																																																																																																																																																																																																																																																																																																																																																																																																																																															
\begin{table}																																																																																																																																																																																																																																																																																																																																																																																																																																																																																																																																																																																																																																																																																																																																																																																																																																																																																																																																																																																																																																																															
\begin{center}																																																																																																																																																																																																																																																																																																																																																																																																																																																																																																																																																																																																																																																																																																																																																																																																																																																																																																																																																																																																																																																															
\caption{Results of Voigt profile fitting for lower ions in the $z=2.5036$ absorber toward J0230-0334}																																																																																																																																																																																																																																																																																																																																																																																																																																																																																																																																																																																																																																																																																																																																																																																																																																																																																																																																																																																																																																																															
\vskip 5pt																																																																																																																																																																																																																																																																																																																																																																																																																																																																																																																																																																																																																																																																																																																																																																																																																																																																																																																																																																																																																																																															
\begin{tabular}{ccccccc}																																																																																																																																																																																																																																																																																																																																																																																																																																																																																																																																																																																																																																																																																																																																																																																																																																																																																																																																																																																																																																																															
\tableline\tableline																																																																																																																																																																																																																																																																																																																																																																																																																																																																																																																																																																																																																																																																																																																																																																																																																																																																																																																																																																																																																																																															
$z$			&	$b^{1}_{eff}$	&log$N_{\rm Mg I}$	&log$N_{\rm Fe II}$	&log$N_{\rm Zn II}$	&log$N_{\rm Ni II}$	&log$N_{\rm Cr II}$	\\
\end{tabular}																																																																																																																																																																																																																																																																																																																																																																																																																																																																																																																																																																																																																																																																																																																																																																																																																																																																																																																																																																																																																																																															
\vskip 5pt																																																																																																																																																																																																																																																																																																																																																																																																																																																																																																																																																																																																																																																																																																																																																																																																																																																																																																																																																																																																																																																															
\begin{tabular}{cccccc}																																																																																																																																																																																																																																																																																																																																																																																																																																																																																																																																																																																																																																																																																																																																																																																																																																																																																																																																																																																																																																																															
\tableline\tableline																																																																																																																																																																																																																																																																																																																																																																																																																																																																																																																																																																																																																																																																																																																																																																																																																																																																																																																																																																																																																																																															
$z$			&	$b^{1}_{eff}$	&	log$N_{\rm Si II}$	&	log$N_{\rm S II}$	&	log$N_{\rm Mn II}$	&	log$N_{\rm Al III}$	\\
\end{tabular}																																																																																																																																																																																																																																																																																																																																																																																																																																																																																																																																																																																																																																																																																																																																																																																																																																																																																																																																																																																																																																																															
\tablenotetext{1}{$b_{eff}$ denotes the effective Doppler $b$ parameter in km s$^{-1}$. Logarithmic column densities are in cm$^{-2}$.}																																																																																																																																																																																																																																																																																																																																																																																																																																																																																																																																																																																																																																																																																																																																																																																																																																																																																																																																																																																																																																																															
\end{center}																																																																																																																																																																																																																																																																																																																																																																																																																																																																																																																																																																																																																																																																																																																																																																																																																																																																																																																																																																																																																																																															
\end{table}																																																																																																																																																																																																																																																																																																																																																																																																																																																																																																																																																																																																																																																																																																																																																																																																																																																																																																																																																																																																																																																															
\end{landscape}																																																																																																																																																																																																																																																																																																																																																																																																																																																																																																																																																																																																																																																																																																																																																																																																																																																																																																																																																																																																																																																															
																																																																																																																																																																																																																																																																																																																																																																																																																																																																																																																																																																																																																																																																																																																																																																																																																																																																																																																																																																																																																																																															

\begin{table}																																																																																																																																																																																																																																																																																																																																																																																																																																																																																																																																																																																																																																																																																																																																																																																																																																																																																																																																																																																																																																																															
\begin{center}																																																																																																																																																																																																																																																																																																																																																																																																																																																																																																																																																																																																																																																																																																																																																																																																																																																																																																																																																																																																																																																															
\caption{Results of Voigt profile fitting for higher ions in the $z=2.5036$ absorber toward J0230-0334}																																																																																																																																																																																																																																																																																																																																																																																																																																																																																																																																																																																																																																																																																																																																																																																																																																																																																																																																																																																																																																																															
\vskip 5pt																																																																																																																																																																																																																																																																																																																																																																																																																																																																																																																																																																																																																																																																																																																																																																																																																																																																																																																																																																																																																																																															
\begin{tabular}{cccc}																																																																																																																																																																																																																																																																																																																																																																																																																																																																																																																																																																																																																																																																																																																																																																																																																																																																																																																																																																																																																																																															
\tableline\tableline																																																																																																																																																																																																																																																																																																																																																																																																																																																																																																																																																																																																																																																																																																																																																																																																																																																																																																																																																																																																																																																															
$z$			&	$b^{1}_{eff}$	&	log$N_{\rm C IV}$	&	log$N_{\rm Si IV}$	\\
\end{tabular}																																																																																																																																																																																																																																																																																																																																																																																																																																																																																																																																																																																																																																																																																																																																																																																																																																																																																																																																																																																																																																																															
\tablenotetext{1}{$b_{eff}$ denotes the effective Doppler $b$ parameter in km s$^{-1}$. Logarithmic column densities are in cm$^{-2}$.}																																																																																																																																																																																																																																																																																																																																																																																																																																																																																																																																																																																																																																																																																																																																																																																																																																																																																																																																																																																																																																																															
\end{center}																																																																																																																																																																																																																																																																																																																																																																																																																																																																																																																																																																																																																																																																																																																																																																																																																																																																																																																																																																																																																																																															
\end{table}

\begin{table}
\begin{center}
\caption{Total column densities for the $z=2.5036$ absorber toward J0230-0334}
\vskip 5pt
\begin{tabular}{ccc}
\tableline\tableline
Ion    &    log $N^{fit}$    &    log $N^{AOD}$    \\
    &    (cm$^{-2}$)    &    (cm$^{-2}$)    \\
\tableline
H I    &    $21.74\pm0.10$    &    ...        \\   
C IV    &    $14.60\pm0.05$    &    $14.59\pm0.01$    \\
B II    &    ...        &    $<11.78$    \\
Mg I    &    $12.83\pm0.12$    &    $12.85\pm0.05$    \\
Al III    &    $13.19\pm0.08$    &    $13.14\pm0.06$    \\
Si II    &    $16.15\pm0.04$    &    $15.96\pm0.02$    \\
Si II*    &    ...        &    $<12.85$    \\
Si IV    &    $13.89\pm0.04$    &    $13.86\pm0.03$    \\
S II    &    $15.80\pm0.11$    &    $15.72\pm0.02$    \\
Ti II    &    ...        &    $<12.83$    \\
V II    &    ...        &    $<12.83$    \\   
Cr II    &    $14.03\pm0.07$    &    ...        \\
Mn II    &    $13.43\pm0.05$    &    $13.36\pm0.03$    \\
Fe II    &    $15.59\pm0.06$    &    $15.64\pm0.03$    \\
Co II    &    ...        &    $<13.23$    \\   
Ni II    &    $14.32\pm0.05$    &    $14.32\pm0.03$    \\   
Cu II    &    ...        &    $<12.33$    \\
Zn II &    $12.95\pm0.09$    &    ...        \\
Ge II    &    ...        &    $<12.82$    \\
\tableline
\end{tabular}
\end{center}
\end{table}


																																																																																																																																																																																																																																																																																																																																																																																																																																																																																																																																																																																																																																																																																																																																																																																																																																																																																																																																																																																																																																																															
\begin{table}																																																																																																																																																																																																																																																																																																																																																																																																																																																																																																																																																																																																																																																																																																																																																																																																																																																																																																																																																																																																																																																															
\begin{center}																																																																																																																																																																																																																																																																																																																																																																																																																																																																																																																																																																																																																																																																																																																																																																																																																																																																																																																																																																																																																																																															
\caption{Measured element abundances relative to solar for the $z=2.5036$ absorber toward J0230-0334}																																																																																																																																																																																																																																																																																																																																																																																																																																																																																																																																																																																																																																																																																																																																																																																																																																																																																																																																																																																																																																																															
\vskip 5pt																																																																																																																																																																																																																																																																																																																																																																																																																																																																																																																																																																																																																																																																																																																																																																																																																																																																																																																																																																																																																																																															
\begin{tabular}{cc}																																																																																																																																																																																																																																																																																																																																																																																																																																																																																																																																																																																																																																																																																																																																																																																																																																																																																																																																																																																																																																																															
\tableline\tableline																																																																																																																																																																																																																																																																																																																																																																																																																																																																																																																																																																																																																																																																																																																																																																																																																																																																																																																																																																																																																																																															
Element	&	[X/H]$^{1}$	\\
\end{tabular}																																																																																																																																																																																																																																																																																																																																																																																																																																																																																																																																																																																																																																																																																																																																																																																																																																																																																																																																																																																																																																																															
\tablenotetext{1}{Abundance estimates based on the dominant metal ionization state and H I.}																																																																																																																																																																																																																																																																																																																																																																																																																																																																																																																																																																																																																																																																																																																																																																																																																																																																																																																																																																																																																																																															
\end{center}																																																																																																																																																																																																																																																																																																																																																																																																																																																																																																																																																																																																																																																																																																																																																																																																																																																																																																																																																																																																																																																															
\end{table}																																																																																																																																																																																																																																																																																																																																																																																																																																																																																																																																																																																																																																																																																																																																																																																																																																																																																																																																																																																																																																																															
\clearpage																																																																																																																																																																																																																																																																																																																																																																																																																																																																																																																																																																																																																																																																																																																																																																																																																																																																																																																																																																																																																																																															
																																																																																																																																																																																																																																																																																																																																																																																																																																																																																																																																																																																																																																																																																																																																																																																																																																																																																																																																																																																																																																																															
\begin{landscape}
\begin{table}
\begin{center}
\caption{Results of Voigt profile fitting for lower ions in the $z=2.045$ absorber toward Q0743+1421}
\vskip 1pt
\begin{tabular}{cccc cccc }																	
\tableline\tableline																	

$ z$					&	$b^{1}_{eff}$		&	$N_{\rm Al II}$	&	$N_{\rm Al III}$		&	$N_{\rm C II}$	&	$N_{\rm C II^{*}}$	&	$N_{\rm Cr II}$		&	$N_{\rm Fe II}$	\\
\tableline															
$2.04464\pm0.000003$	&	$17.799\pm0.378$	&	$>14.671$	&	$12.933\pm0.0326$	&	$>16.906$	&	$>14.961$		&	$13.723\pm0.0213$	&	$15.403\pm0.0449$	\\
\tableline															
\end{tabular}															
\vskip 5pt															
\begin{tabular}{cccc cccc}															
$ z$					&	$b^{1}_{eff}$		&	$N_{\rm Mg I}$		&	$N_{\rm Mg II}$	&	$N_{\rm Mn II}$		&	$N_{\rm Ni II}$		&	$N_{\rm Si II}$	&	$N_{\rm Zn II}$	\\
\tableline															
$2.04464\pm0.000003$	&	$17.799\pm0.378$	&	$12.968\pm0.0385$	&	$>15.590$	&	$13.180\pm0.0454$	&	$14.123\pm0.0300$	&	$>16.083$	&	$13.048\pm0.0268$	\\

\tableline																	
\tableline																	
\end{tabular}																	
\tablenotetext{1}{$b_{eff}$ denotes the effective Doppler $b$ parameter in km s$^{-1}$. Logarithmic column densities are in cm$^{-2}$.}
\end{center}																	
\end{table}																	
\end{landscape}

\begin{landscape}												
\begin{table}												
\begin{center}												
\caption{Results of Voigt profile fitting for higher ions in the $z=2.045$ sub-DLA toward Q0743+1421}												
\vskip 15pt												
\begin{tabular}{cccc}												
\tableline\tableline												
$ z$					&	$b^{1}_{eff}$		&	$N_{\rm C IV}$	&	$N_{\rm Si IV}$	\\
\tableline							
$2.0440\pm0.000016$	&	$26.432\pm2.740$	&	$14.407\pm0.0708$	&	$13.845\pm0.0574$	\\
\tableline												
\end{tabular}
\tablenotetext{1}{$b_{eff}$ denotes the effective Doppler $b$ parameter in km s$^{-1}$. Logarithmic column densities are in cm$^{-2}$.}
\end{center}												
\end{table}												
\end{landscape}

\begin{table}	
							
\begin{center}								
\caption{Total Column Densities for the $z=2.045$ Absorber toward Q0743+1421}								
\vskip 10pt								
\begin{tabular}{ccc}								
\tableline\tableline								
Ion	&	log $N^{fit}$		&	log $N^{AOD}$		\\			
	&	(cm$^{-2}$)		&	(cm$^{-2}$)		\\			
\tableline											

H I	&	$21.90\pm0.10$		&	\ldots			\\	
C I	&	\ldots			&	$<13.08$		\\				
C II	&	$>16.91$		&	$>14.82$		\\	
C II*	&	$>14.96$		&	$>14.46$		\\	
C IV	&	$14.41\pm0.07$	&	$14.21\pm0.03$	\\	
B II	&	\ldots			&	$<12.26$		\\	
Mg I	&	$12.97\pm0.04$	&	$12.71\pm0.02$	\\	
Mg II	&	$>15.59$		&	$>13.57$		\\	
Al II	&	$>14.67$		&	$>13.35$		\\	
Al III	&	$12.93\pm0.03$	&	$12.82\pm0.04$	\\					
Si II	&	$>16.08$		&	$>15.76$		\\	
Si II*	&	\ldots			&	$<13.01$		\\	
Si IV	&	$13.84\pm0.06$	&	$13.76\pm0.06$	\\	
V II	&	\ldots			&	$<12.79$		\\				
Cr II	&	$13.72\pm0.02$	&	$13.62\pm0.02$	\\
Mn II	&	$13.18\pm0.05$	&	$13.04\pm0.03$	\\	
Fe II	&	$15.40\pm0.04$	&	$15.46\pm0.06$	\\			
Co II	&	\ldots			&	$<13.67$		\\	
Ni II	&	$14.12\pm0.03$	&	$14.20\pm0.04$	\\				
CuII	&	\ldots			&	$<12.90$		\\				
Zn II	&	$13.05\pm0.03$	&	\ldots			\\			
Ga II	&	\ldots			&	$<12.07$		\\				
Ge II	&	\ldots			&	$<12.88$		\\

\tableline								
\end{tabular}								
\end{center}								
\end{table}

\begin{table}									
\begin{center}									
\caption{Measured Element Abundances Relative to Solar for the $z=2.045$ Absorber toward Q0743+1421}									
\vskip 10pt									
\begin{tabular}{cc}									
\tableline\tableline																	
Element	&	[X/H]$^{1}$\\
\tableline									

C	&	$>-3.51$	\\
B	&	$<-0.3$	\\
Mg	&	$>-3.93$	\\
Al	&	$>-3.00$	\\
Si	&	$>-1.65$	\\
V	&	$<-1.0$	\\
Cr	&	$-1.82\pm0.10$	\\
Mn	&	$-2.15\pm0.11$	\\
Fe	&	$-2.00\pm0.11$	\\
Co	&	\ldots	\\
Ni	&	$-2.00\pm0.10$	\\
Cu	&	$<-1.2$	\\
Zn	&	$-1.41\pm0.10$	\\
Ga	&	$<-0.9$	\\
Ge	&	$<-0.7$	\\

\tableline									
\end{tabular}									
\tablenotetext{1}{Abundance estimates based on the dominant metal ionization state and H I.}									
\end{center}									
\end{table}									
\clearpage									

\begin{table}
\caption{Results of Voigt profile fitting for low ions in the $z=2.392$ absorber toward Q1418+0718}
\vskip 5pt
\begin{tabular}{ccccc}
\tableline\tableline
$z$ 			& 	$b^{1}_{eff}$	&	log$N_{\rm C II*}$	& log$N_{\rm O I}$ & log $N_{\rm Si II}$ \\

\tableline
$2.39183\pm0.000004$	&	$8.36 \pm 0.27$	& $13.65\pm0.15$&$17.72 \pm 0.12$& $15.94 \pm 0.07$\\
$2.39217\pm0.000006$	&	$6.24 \pm 0.54$	&...&	$14.53\pm0.19$&$14.77\pm0.13$\\
\tableline															
\end{tabular}		

\vskip 5pt			
\begin{tabular}{cccccc}
\tableline
$z$ 			& 	$b^{1}_{eff}$	& log $N_{\rm S II}$ & log $N_{\rm Ti II}$ & log $N_{\rm Cr II}$ & log $N_{\rm Fe II}$\\
\tableline
$2.39183\pm0.000004$	&	$8.36 \pm 0.27$&$15.84 \pm 0.16$& $12.65 \pm 0.23$& $13.68 \pm 0.07$& $15.21 \pm 0.06$\\
$2.39217\pm0.000006$	&	$6.24 \pm 0.54$&...&...&$12.87 \pm 0.25$&$14.25 \pm 0.07$	\\
\tableline															
\end{tabular}			

\vskip 5pt															
\begin{tabular}{cccccc}	
\tableline
$z$ 			& 	$b^{1}_{eff}$	 &  log $N_{\rm Ni II}$ &log $N_{\rm Mn II}$ & log $N_{\rm Zn II}$ \\
\tableline
$2.39183\pm0.000004$	&	$8.36 \pm 0.27$& $13.77 \pm 0.06$ & $12.85 \pm 0.08$&$12.79\pm0.21$		\\
$2.39217\pm0.000006$	&	$6.24 \pm 0.54$& $13.15\pm0.16$&...&...	\\
\tableline
\end{tabular}
\tablenotetext{1}{$b_{eff}$ denotes the effective Doppler $b$ parameter in km s$^{-1}$. Logarithmic column densities are in cm$^{-2}$.}
\end{table}

\begin{table}								
\begin{center}								
\caption{Total Column Densities for the $z=2.392$ Absorber toward Q1418+0718}								
\vskip 10pt								
\begin{tabular}{ccc}								
\tableline\tableline								
Ion	&	log $N^{fit}$	&	log $N^{AOD}$	\\			
	&	(cm$^{-2}$)	&	(cm$^{-2}$)	\\			
\tableline
H I	&	$21.70\pm0.10$		&	\ldots			\\	
B II   &      \ldots & $< 12.06$  \\
C II	&	...		&	$>14.84$		\\	
C II*	&	$13.65 \pm 0.15$		&	$13.66 \pm 0.15$		\\	
O I    &      $\ge 17.72$                    &  $> 15.13$\\	
Si II	&	$15.97 \pm 0.07$		&	$15.61 \pm 0.11$		\\	
Si II* &     \ldots &$<12.00$\\
Si IV &      ...& $13.20 \pm 0.17$\\
S II  &        $15.84\pm 0.16$ & $> 15.17$\\
Ti II	&	$12.65 \pm 0.23$			&$12.96 \pm 0.19$				\\	
V II  &         \ldots &$< 12.55$  \\
Cr II	&	$13.74\pm0.07$	&	$13.79 \pm 0.10$	\\	
Mn II	&	$12.85\pm0.08$	&	$12.83 \pm 0.12$	\\	
Fe II	&	$15.25\pm0.05$	&	$>14.82$	\\	
Co II &       \ldots &$<13.07$\\
Ni II	&	$13.86\pm0.06$	&	$13.93 \pm 0.15$	\\	
Cu II   &     \ldots & $<12.98$\\
Zn II	&	$12.79\pm0.21$	&	$12.93 \pm 0.15$			\\	

\tableline								
\end{tabular}								
\end{center}								
\end{table}			

\begin{table}									
\begin{center}									
\caption{Measured Element Abundances Relative to Solar for the $z=2.392$ absorber toward Q1418+0718}									
\vskip 10pt									
\begin{tabular}{cc}									
\tableline\tableline																	
Element	&	[X/H]$^{1}$\\
\tableline									
C	&	$\ge -0.49$	\\
B     &        $<-0.34$ \\
O     &       $\ge -0.67$ \\
Si	&	$-1.24\pm0.12$	\\
S      &       $-0.98\pm 0.19$        \\
Ti      &      $-2.00\pm0.29$         \\
 V      &     $<-1.34$\\
Cr	&	$-1.60\pm0.12$	\\
Mn	&	$-2.28\pm0.13$	\\
Fe	&	$-1.95\pm0.11$	\\
Co    &       $<-1.62$\\
Ni	&	$-2.06\pm0.11$	\\
Cu    &       $<-1.15$\\
Zn	&	$-1.47\pm0.25$	\\
\tableline									
\end{tabular}									
\tablenotetext{1}{Abundance estimates based on the dominant metal ionization state and H I.}									
\end{center}									
\end{table}																	
\clearpage

\begin{table}									
\begin{center}									
\caption{Summary of Depletion vs. Metallicity Statistics}									
\vskip 10pt									
\begin{tabular}{cccc}									
\tableline\tableline																	
Sample,	&	Statistic & [Fe/X] vs. [X/H] & [Si/X] vs. [X/H]]\\
$N_{\rm H I}$ Range&&&\\
\tableline
Moderate DLAs & Bisector Fit Slope & -0.500 $\pm$ 0.043  &  -0.406 $\pm$ 0.081\\
$(20.3 \le$log $N_{\rm H I}$$< 21.7)$&Bisector Fit Intercept& -1.06$\pm$0.059& -0.514 $\pm$ 0.091\\
Super-DLAs &  Bisector Fit Slope &  -0.774 $\pm$ 0.170 &  -0.551 $\pm$ 0.184 \\
(log$N_{\rm H I} \ge 21.7$)& Bisector Fit Intercept & -1.52 $\pm$ 0.24 & -0.677$\pm$0.208\\
\tableline					
Moderate DLAs & Spearman $r_{S}$, $P$ & -0.596, $<$ 1.0$ \times$ 10$^{-6}$ &  -0.404, 1.46 $\times$ 10$^{-3}$\\
($20.3 \le$log $N_{\rm H I}$$< 21.7$)&&&\\
Super-DLAs, & Spearman  $r_{S}$, $P$ &  -0.441, 6.57 $\times$ 10$^{-2}$& -0.878, $<$ 5.0$ \times$ 10$^{-2}$\\
(log$N_{\rm H I} \ge 21.7$)&&&\\				
\tableline	
								
\end{tabular}																
\end{center}									
\end{table}																	
\clearpage						

\end{document}